\documentclass[letterpaper]{JHEP3} 

\newcommand{\be}{\begin{equation}}
\newcommand{\ee}{\end{equation}}
\newcommand{\bear}{\begin{eqnarray}}
\newcommand{\eear}{\end{eqnarray}} \newcommand{\ba}{\begin{array}}
\newcommand{\ea}{\end{array}}

\newcommand{\LL}{L}
\newcommand{\vb}{\right|}
\newcommand{\CL}{{\cal L}} 

\newcommand{\Ztwol}{\ensuremath{Z^{\rm AP}_2}}

\title{\LARGE \bf Chiral Compactification on a Square}
\author{Bogdan A.~Dobrescu, Eduardo Pont\'{o}n\\
Fermi National Accelerator Laboratory, Batavia, IL 60510 \\ 
\email{bdob@fnal.gov, eponton@fnal.gov}}
\abstract{
We study quantum field theory in six dimensions with two of them
compactified on a square. A simple boundary condition
is the identification of two pairs of adjacent sides of the square
such that the values of a field at two identified points differ
by an arbitrary phase. This allows
a chiral fermion content for the four-dimensional theory obtained
after integrating over the square. 
We find that nontrivial solutions for the field equations exist 
only when the phase is a multiple of $\pi/2$, so that 
this compactification turns out to be equivalent to a $T^2/Z_4$ 
orbifold associated with toroidal boundary conditions that are 
either periodic or anti-periodic.
The equality of the Lagrangian densities at the identified points 
in conjunction with six-dimensional Lorentz invariance leads 
to an exact $Z_8\times Z_2$ symmetry, where 
the $Z_2$ parity ensures the stability of the
lightest Kaluza-Klein particle.
}

\preprint{hep-th/0401032 \\ FERMILAB-Pub-03/414-T \\ 
January 6, 2004 \ (Revised September 15, 2004)} 
\keywords{extra dimensions; field-theory orbifolds; Kaluza-Klein wave functions}

\begin{document}


\section{Introduction} \setcounter{equation}{0}

Quantum field theory in six dimensions has been studied in connection
to various alternatives for physics beyond the standard model, and was
shown to provide explanations for proton stability
\cite{Appelquist:2001mj}, the origin of electroweak symmetry breaking
\cite{Arkani-Hamed:2000hv, Hashimoto:2000uk, Csaki:2002ur, Scrucca:2003ut}, the number
of fermion generations \cite{Dobrescu:2001ae, Fabbrichesi:2001fx,
Borghini:2001sa, Fabbrichesi:2002am, Frere:2001ug, Watari:2002tf}, 
and the breaking of grand unified gauge groups \cite{Hebecker:2001jb,
Hall:2001xr, Asaka:2002my}.

An important question is how are the two extra dimensions
compactified.  If any of the standard model fermions propagate in the
extra dimensions, a restriction on the compactification comes from the
requirement of having chiral fermions in the four-dimensional
low-energy theory.  Only few examples of compactifications with this
property have been analyzed in detail so far.  The $T^2/Z_2$ orbifold is a a
parallelogram folded once onto itself (see, {\it
e.g.,}~\cite{Ponton:2001hq}).  The field decomposition in Kaluza-Klein (KK)
modes is given in \cite{Appelquist:2000nn} for the case where the
parallelogram is a rectangle.  The $T^2/Z_4$ orbifold, which is a
compactification on a square, has the merit of automatically
preserving a $Z_8$
subgroup of the six-dimensional Lorentz symmetry
\cite{Appelquist:2001mj} which ensures a long proton lifetime, and
forces the neutrino masses to be of the Dirac type
\cite{Appelquist:2002ft}.  
In a different context, the $T^2/Z_4$ orbifold was shown to allow 
the Higgs doublet be part of a $G_2$ gauge field \cite{Csaki:2002ur}.
A discussion of $T^2/Z_N$ orbifolds in general 
can be found in \cite{Scrucca:2003ut} while more general manifolds
with conical singularities were considered in \cite{Hebecker:2003jt}.

In the case of five-dimensional theories, any orbifold compactification is
equivalent to a set of chiral boundary conditions (see, {\it
e.g.,}~\cite{Cheng:1999bg}), while the reverse is not true
\cite{Csaki:2003dt, Csaki:2003sh}.
For six-dimensional theories, however, neither 
the boundary conditions associated
with various orbifolds nor the
general restrictions imposed by the action principle on the 
boundary conditions have been presented in the literature.

In this paper we explore a chiral compactification on a square.  We
derive a simple set of boundary conditions, and prove that it 
allows nontrivial solutions to the field equations only when it
is equivalent to the $T^2/Z_4$ orbifold.
We also determine the KK wave
functions for scalars and fermions, and identify the unbroken 
subgroup of the six-dimensional Lorentz symmetry. 

The boundary conditions and the solutions to the field equations
for free scalar and fermion fields are analyzed in Sections 2 and 3.
Section 4 is devoted to the symmetries of the 
action. Interactions among KK modes are studied in Section 5,
while the connection to orbifolds is presented in Section 6.
A summary of results is given in Section 7.
 
\section{Compactification on a square}
\setcounter{equation}{0}

The spacetime considered in this paper is six-dimensional: four
spacetime dimensions of coordinates $x^\mu$, $\mu =0,1,2,3$, form the
usual Minkowski spacetime, and two transverse spatial dimensions of
coordinates $x^4$ and $x^5$ are flat and compact.  We analyze a simple
compactification: a square with $0 \leq x^4, x^5 \leq \LL$.

In this section we study the behavior of free scalar fields,
$\Phi(x^\mu, x^4, x^5)$, described by the following action:
\be
S_\Phi = \int d^4 x \int_0^{\LL} dx^4 \int_0^{\LL} dx^5
\left( \partial_\alpha \Phi^\dagger \partial^\alpha \Phi 
- M_0^2 \Phi^\dagger\Phi\right) ~.
\label{action}
\ee
We use letters from the beginning of the Greek alphabet to label the
six-dimensional coordinates $\alpha, \beta, ...  = 0,1, ..., 5$, and
letters from the middle of the Greek alphabet to label the Minkowski
coordinates $\mu, \nu, ...  = 0,1, 2, 3$.

Under a variation of the field, $\delta\Phi(x^\mu, x^4, x^5)$, the
variation of the action is given by
\be
\delta S_\Phi  = \delta S_\Phi^v  + \delta S_\Phi^s ~,
\ee
where the first term is a ``volume'' integral,
\be
\delta S_\Phi^v = -  \int d^4 x 
\int_0^{\LL} dx^4 \int_0^{\LL} dx^5
\left( 
\partial^\alpha \partial_\alpha \Phi^\dagger
- M_0^2 \Phi^\dagger \right) \delta\Phi  ~,
\ee
and the second term is a ``surface'' integral,
\bear
\delta S_\Phi^s & = &  \int d^4 x \left[
\int_0^{\LL} dx^4 
\left(\left.\partial_5 \Phi^\dagger \delta\Phi \vb_{x^5=\LL}
- \left.\partial_5 \Phi^\dagger \delta\Phi \vb_{x^5=0} \right)\right.
\nonumber \\ [.3em] 
& & + \left.\int_0^{\LL} dx^5 
\left(\left.\partial_4 \Phi^\dagger \delta\Phi \vb_{x^4=\LL}
- \left.\partial_4 \Phi^\dagger \delta\Phi \vb_{x^4=0} \right) 
\right] ~.
\label{surface}
\eear
Here we have assumed as usual that the field vanishes at $x^\mu
\rightarrow \pm\infty$.  

Given that the action has to be stationary
with respect to {\it any} variation of the field, the volume and
surface terms must vanish independently.  Requiring $\delta S_\Phi^v =
0$ implies that $\Phi$ is a solution to the six-dimensional
Klein-Gordon equation, while $\delta S_\Phi^s = 0$ forces the 
boundary conditions that can be
imposed on $\Phi$ to obey a certain restriction. To derive it, 
we rewrite the surface integral as
\bear
\delta S_\Phi^s & = &  \int d^4 x 
\int_0^{\LL} dy
\left(
\left.\partial_4 \Phi^\dagger
\delta\Phi \vb_{(x^4,x^5)=(\LL, y)}
+ \left.\partial_5 \Phi^\dagger
\delta\Phi \vb_{(x^4,x^5)=(y,\LL)}
\right.
\nonumber \\ [.6em] 
& &  \hspace*{6em} 
\left.
- \left.\partial_4 \Phi^\dagger
\delta\Phi \vb_{(x^4,x^5)=(0, y)} 
- \left.\partial_5 \Phi^\dagger
\delta\Phi \vb_{(x^4,x^5)=(y, 0)} 
\right) = 0 
 ~.
\label{surface-second}
\eear
The requirement that the action be stationary for 
{\it arbitrary} variations of the field satisfying certain 
boundary conditions implies that the integrand vanishes point by point: 
\bear
\label{generalBC}
\left.\partial_4 \Phi^\dagger \delta\Phi \vb_{(x^4,x^5)=(\LL, y)}
+ \left.\partial_5 \Phi^\dagger \delta\Phi \vb_{(x^4,x^5)=(y,\LL)}
= \left.\partial_4 \Phi^\dagger \delta\Phi \vb_{(x^4,x^5)=(0, y)} 
+ \left.\partial_5 \Phi^\dagger \delta\Phi \vb_{(x^4,x^5)=(y, 0)} 
\nonumber \\ [.6em] {} \\ [-2em] {} \nonumber 
\eear
for any $y \in [0,\LL]$.

Even with this restriction, there are many possible boundary
conditions.  Identifying the opposite sides of the square, which
produces a torus, is an obvious example.  However, it is well known
that the toroidal compactification does not allow chiral fermions in
the four-dimensional effective theory.  A particular class of boundary
conditions which allows four-dimensional chiral fermions (as shown
later, in Section 3) is presented next.

\FIGURE{
\begin{picture}(167,170)(-33,-35)
\multiput(-0.3,-0.3)(0, 2.62){28}{\line(1,0){70}}
\multiput(-0.3,-0.3)(2.62, 0){28}{\line(0,1){70}}
\put(0,-20){\vector(0, 1){136}}
\put(-20,0){\vector(1, 0){150}}
\thicklines
\put(0,0){\line(1, 0){70}}
\put(0,1){\line(1, 0){70}}
\put(0,-1){\line(1, 0){70}}
\put(0,0){\line(0, 1){70}}
\put(-1,0){\line(0, 1){70}}
\put(1,0){\line(0, 1){70}}
\put(70,70){\line(-1, 0){70}}
\put(70,70){\line(0, -1){70}}
\put(127,-15){$x^4$}
\put(-20,113){$x^5$}
\put(-12,-12){$0$}
\put(-19,65){$ \LL$}
\put(65,-15){$\LL$}
\put(.5,.5){\circle*{7}}
\put(69,69){\circle*{7}}
\end{picture}
\vspace*{-0.8em}
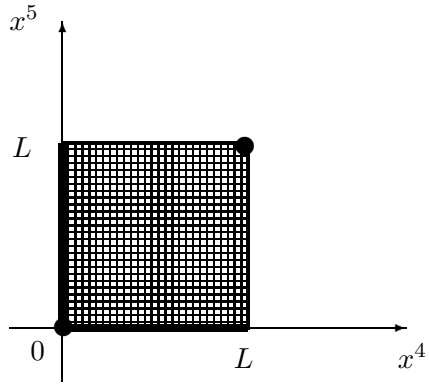
\hspace*{-0.5em}\caption{Square compactification with identified
pairs of adjacent sides. \\ } 
\label{appfig} 
}
Consider the identification of two pairs of adjacent sides of the
square:
\be
(y, 0 ) \equiv (0, y) \; , \;\; (y, \LL) \equiv (\LL, y) \; , \;\; 
\forall y\in [0, \LL] ~.
\label{fold}
\ee
Topologically, this is equivalent to folding the square along a
diagonal and gluing the boundary.  Note though that we choose the
metric on the square to be flat.  There are two points (see figure 1)
that remain invariant under the folding operation (\ref{fold}):
$(0,0)$ and $(\LL, \LL)$.

We interpret the identification of different sides of the square as
the requirement that the Lagrangian density at points identified by
the folding operation (\ref{fold}) is the same:
\bear
\CL(x^\mu, y,0) = \CL(x^\mu, 0, y) ~,
\nonumber \\ [.3em] 
\CL(x^\mu, y,\LL) = \CL(x^\mu, \LL, y) ~,
\eear
where $\CL$ is the integrand shown in Eq.~(\ref{action}).  This
ensures that the physics at identified points is the same.  On the
other hand, the field at two identified points does not need to be the
same.  The global $U(1)$ symmetry of the Lagrangian suggests that the
field values at two identified points may differ by a constant phase:
\be
\Phi(x^\mu, y, 0)  =  e^{i \theta} \Phi (x^\mu, 0, y) ~,
\label{phase}
\ee
for any $y \in [0,\LL]$.  Taking the derivative with respect to $y$ in
the above equations we find that
\be
\left.\partial_4\Phi \vb_{(x^4, x^5)=(y,0)} = e^{i \theta}
\left.\partial_5\Phi \vb_{(x^4, x^5)=(0,y)} ~.
\label{auto-der-phase}
\ee
Using these conditions we obtain
\be
\CL(x^\mu, 0, y) - \CL(x^\mu, y,0) = 
\left.\partial_5\Phi^\dagger \partial_5\Phi \vb_{(x^4, x^5)=(y,0)}
- \left.\partial_4\Phi^\dagger \partial_4\Phi \vb_{(x^4, x^5)=(0,y)} 
~.
\ee
Therefore, the Lagrangian is the same at two identified points only
if, in addition to Eq.~(\ref{phase}), the following condition on the
derivatives is satisfied:
\be
\left.\partial_5\Phi \vb_{(x^4, x^5)=(y,0)}  =  e^{i \theta^\prime} 
\left.\partial_4\Phi \vb_{(x^4, x^5)=(0,y)} ~.
\label{der-phase}
\ee
This has a simple geometrical interpretation: the folding of the
square is smooth, {\it i.e.}, the derivative of the field in the
direction perpendicular to the identified boundaries is continuous
up to a phase.

Conditions analogous to (\ref{phase}) and (\ref{der-phase}) have to be
imposed on the other two sides of the square, but the phases may be
different both for the fields,
\be
\Phi(x^\mu, y, \LL) = e^{i \tilde{\theta}} \Phi (x^\mu, \LL, y) ~,
\label{other-phase}
\ee
and for the derivatives,
\be
\left.\partial_5\Phi \vb_{(x^4, x^5)=(y,\LL)}  = 
e^{i \tilde{\theta}^\prime} 
\left.\partial_4\Phi \vb_{(x^4, x^5)=(\LL,y)} ~.
\label{other-der-phase}
\ee

Although the phases $\theta$, $\tilde{\theta}$, $\theta^\prime$ and
$\tilde{\theta}^\prime$ may be different in general, there are certain
constraints on them.  Most importantly, the boundary conditions must
be consistent with the stationarity of the action, {\it i.e.}, the
general condition (\ref{generalBC}).  Given that the variation
$\delta\Phi$ on the boundary has to obey the same constraints as
$\Phi$, namely Eqs.~(\ref{phase}) and (\ref{other-phase}), we can use
Eqs.~(\ref{der-phase}) and (\ref{other-der-phase}) to write
Eq.~(\ref{generalBC}) as
\be
\left[e^{i (\tilde{\theta}^\prime - \tilde{\theta})} +1 \right]
\left. \partial_5 \Phi^\dagger \delta\Phi \vb_{(x^4,x^5)=(y,\LL)}
= \left[e^{i (\theta^\prime - \theta)} +1 \right]
\left. \partial_5 \Phi^\dagger \delta\Phi \vb_{(x^4,x^5)=(y,0)}  ~.
\label{surface-fold}
\ee
This condition must be satisfied for any field variation $\delta\Phi$,
so that the left- and right-handed sides of Eq.~(\ref{surface-fold})
must vanish independently. Hence, two constraints can be derived:
\be
e^{i\theta^\prime} = - e^{i\theta} \;  , \; {\rm or}  \;  \;
\delta\Phi\left.\partial_5 \Phi\vb_{(x^4,x^5)=(y,0)} = 0 ~,
\label{prime}
\ee
and 
\be
e^{i\tilde{\theta}^\prime} = - e^{i\tilde{\theta}}
\; , \; {\rm or} \; \; 
\delta\Phi
\left.\partial_5 \Phi\vb_{(x^4,x^5)=(y,L)} = 0 ~.
\label{tilde-prime}
\ee

We now solve the six-dimensional Klein-Gordon equation, \be \left(
\partial^\mu \partial_\mu - \partial_4^2 - \partial_5^2 + M_0^2
\right)\Phi = 0 ~,\ee subject to the ``folding boundary conditions''
(\ref{phase}), (\ref{der-phase}), (\ref{other-phase}),
(\ref{other-der-phase}), with the restrictions (\ref{prime}) and
(\ref{tilde-prime}).  Since the boundary conditions are independent of
$x^\mu$, then $\Phi$ can be decomposed in Fourier modes as follows:
\be
\Phi(x^\mu, x^4, x^5) = \frac{1}{L}
\sum_{j,k} \Phi^{(j,k)} (x^\mu) f^{(j,k)}(x^4, x^5) ~.
\label{fourier}
\ee
The four-dimensional scalar fields $\Phi^{(j,k)}$  satisfy
\be
\left( \partial^\mu \partial_\mu + M_0^2 + M_{j,k}^2 \right)
\Phi^{(j,k)}(x^\mu) = 0 ~,
\ee
where $M_{j,k}^2$ is a positive eigenvalue.  The $f^{(j,k)}$ functions
are solutions to the two-dimensional equation,
\be
\left( \partial_4^2 + \partial_5^2 + M_{j,k}^2 \right)
f^{(j,k)}(x^4, x^5) = 0 ~.
\label{twoD}
\ee
A general solution to the above equation is a linear combination of
eight position-dependent phases,
\bear
\label{general-solution}
f^{(j,k)} = && \hspace*{-.1em}
C_1^+ e^{ i \left(j x^4 + k x^5 \right)/R }
+ C_1^- e^{ - i \left(j x^4 + k x^5 \right)/R}
+ C_2^+ e^{ i \left(j x^4 - k x^5 \right)/R}
+ C_2^- e^{- i \left(j x^4 - k x^5 \right)/R}
\nonumber \\ [.3em]
 + && \hspace*{-.1em} C_3^+ e^{ i \left(k x^4 + j x^5 \right)/R}
+ C_3^- e^{ - i \left(k x^4 + j x^5 \right)/R}
+ C_4^+ e^{ i \left(k x^4 - j x^5 \right)/R}
+ C_4^- e^{- i \left(k x^4 - j x^5 \right)/R} ~, \nonumber\\[-.3em]
\eear
where $j$ and $k$ are real numbers such that
\be
M_{j,k}^2 = \frac{j^2 + k^2}{R^2} ~,
\label{mjk}
\ee
and we defined the ``compactification radius''
\be
R \equiv \frac{\LL}{\pi} ~.
\ee

The eight unknown coefficients, $C^\pm_i$, $i = 1,2,3,4$, that appear
in the general solution, are constrained by the folding boundary
conditions.  Eq.~(\ref{phase}), which relates the field values on the
$x^4=0$ and $x^5=0$ sides of the square, is satisfied for arbitrary 
$j$
and $k$ if and only if
\bear
&& C_3^\pm + C_4^\mp = e^{-i\theta} \left( C_1^\pm + C_2^\pm \right)  
~,
 \nonumber \\ [.3em]
&& C_3^\pm + C_4^\pm = e^{i\theta} \left( C_1^\pm + C_2^\mp \right) ~.
\label{first-set}
\eear
The boundary condition (\ref{der-phase}) that relates the field
derivatives at $x^4=0$ and $x^5=0$ is satisfied for arbitrary $j$ and
$k$ provided
\bear
&& C_3^\pm - C_4^\mp = e^{-i\theta^\prime} \left( C_1^\pm - C_2^\pm 
\right) ~,
\nonumber  \\ [.3em]
&& C_3^\pm - C_4^\pm = e^{i\theta^\prime} \left( C_1^\pm - C_2^\mp 
\right) ~.
\label{second-set}
\eear
For $j=\pm k$, these eight equations are replaced by only four linear
combinations of them, but in the end no new solution is allowed.

The set of eight equations (\ref{first-set}) and (\ref{second-set})
has to be solved subject to the constraint (\ref{prime}).  For
\be
e^{i\theta^\prime} = -e^{i\theta} ~,
\label{phase-relation}
\ee
Eqs.~(\ref{first-set}) and (\ref{second-set}) have nontrivial
solutions only if
\be
e^{4i\theta} = 1 ~.
\label{theta-restriction}
\ee
This is an important restriction on the phase that relates the field
values on the $x^4=0$ and $x^5=0$ boundaries.  Six of the unknown
coefficients are determined in terms of the remaining two, chosen to 
be
$C_{1,2}^+$:
\bear
&& C_3^\pm = C_2^+ e^{\mp i\theta} ~,
\nonumber  \\ [.3em]
&& C_4^\pm = C_1^+ e^{\pm i\theta} ~,
\nonumber  \\ [.3em]
&& C_{1,2}^- = C_{1,2}^+ e^{2 i\theta} ~.
\label{twoparam} 
\eear
If Eq.~(\ref{phase-relation}) is not satisfied, then 
the constraint (\ref{prime}) implies that
$\delta\Phi$ or $\partial_5 \Phi$ vanish at 
$(y,0)$, so that
$f^{(j,k)}$ or $\partial_5 f^{(j,k)}$
vanish at that point.
Eqs.~(\ref{first-set}) and (\ref{second-set}) then have nontrivial
solutions only if 
Eq.~(\ref{twoparam}) is satisfied with 
$e^{2i\theta} = 1$ and $C_2^+ = \mp C_1^+$ 
[the sign is $-$ or $+$ depending on whether $\Phi$
or $\partial_5 \Phi$ vanish at $(y,0)$], 
so that the solutions in this case are subsets of the solutions allowed
by Eq.~(\ref{phase-relation}).  Thus, the most general solution
to the two-dimensional equation (\ref{twoD}) subject to the boundary
conditions (\ref{phase}) and (\ref{der-phase}) is given by
\bear
f^{(j,k)}(x^4, x^5) & = &
2 C_1^+ \left[ e^{-i\theta} 
\cos\left(\frac{j x^4 + k x^5}{R} + \theta \right)
+ \cos \left( \frac{k x^4 - j x^5}{R} + \theta  \right) \right] 
 \nonumber \\ [.3em]
& + & 2 C_2^+  \left[ e^{-i\theta} 
\cos\left( \frac{j x^4 - k x^5}{R} + \theta \right)
+ \cos \left( \frac{k x^4 + j x^5}{R} - \theta  \right) \right] ~.
\label{intermediate-solution}
\eear

Next, we impose the other two boundary conditions, which relate the
$x^4=L$ and $x^5=L$ boundaries, and the constraint
(\ref{tilde-prime}).  Applying the boundary condition for $\Phi$,
Eq.~(\ref{other-phase}), to the solution
(\ref{intermediate-solution}), we find an equation that has to hold
for any $y\in [0,L]$ and arbitrary $j$ and $k$.  Therefore, the
coefficients of $\cos ky/R$, $\sin ky/R$, $\cos jy/R$ and $\sin jy/R$
must vanish independently:
\bear 
&& \left(C_1^+ + C_2^+ \right) 
\left( e^{i ( \tilde{\theta} + \theta )} - 1 \right)
\cos (j\pi + \theta) = 0 ~,
\nonumber  \\ [.3em]
&& \left(C_1^+ - C_2^+ \right) 
\left( e^{i ( \tilde{\theta} + \theta )} + 1 \right)
\sin (j\pi + \theta) = 0 ~,
\nonumber  \\ [.3em]
&& \left(C_1^+ + e^{2i\theta} C_2^+ \right)  
\left( e^{i ( \tilde{\theta} + \theta )} - 1 \right)
\cos (k\pi + \theta) = 0 ~,
\nonumber  \\ [.3em]
&& \left(C_1^+ - e^{2i\theta} C_2^+ \right) 
\left( e^{i ( \tilde{\theta} + \theta )} + 1 \right)
\sin (k\pi + \theta) = 0 ~.
\label{first-four}
\eear
Here we have used the restriction on $\theta$,
Eq.~(\ref{theta-restriction}).  Following the same procedure, the
boundary condition for the derivatives of $\Phi$,
Eq.~(\ref{other-der-phase}), leads to four more equations which can be
obtained from Eqs.~(\ref{first-four}) by substituting
$\tilde{\theta}^\prime$ for $\tilde{\theta}$ and interchanging $\cos$
and $\sin$.
Thus, there are eight equations altogether.  We have to find a
solution to this set of equations with at least one of $C_1^+$ and
$C_2^+$ being nonzero, and which is subject to the constraint
(\ref{tilde-prime}).  This is possible only if $\tilde{\theta}$,
$\tilde{\theta}^\prime$, $j$ and $k$ satisfy certain conditions.  For
\be
e^{i\tilde{\theta}^\prime} = - e^{i\tilde{\theta}} 
\label{other-phase-relation}
\ee
we obtain that either 
\be
e^{i\tilde{\theta}} = e^{i\theta} \; {\rm and} \; j,k \in {\bf Z} ~,
\label{normal-case}
\ee
or 
\be
e^{i\tilde{\theta}} = -e^{i\theta} \; 
{\rm and} \; j+\frac{1}{2}\, , \; k+\frac{1}{2} \in {\bf Z} ~.
\ee
For $f(y,L)\left.\partial_5 f\vb_{(x^4,x^5)=(y,L)} = 0$, the solutions
are again just a subset of the solutions obtained when
Eq.~(\ref{other-phase-relation}) is satisfied, with $e^{2i\theta} = 1$, and
$C_2^+ = -C_1^+$ for $f(y,L)=0$ or $C_2^+ = C_1^+$ for 
$\left.\partial_5 f\vb_{(x^4,x^5)=(y,L)} = 0$.

The conclusion so far is that the most general folding boundary
conditions that allow a nontrivial solution to the six-dimensional
Klein-Gordon equation are given by
\bear
&& \Phi(x^\mu, y, 0) = e^{i n\pi/2} \Phi (x^\mu, 0, y) ~,
\nonumber  \\ [.3em]
&& \Phi(x^\mu, y, \LL) = (-1)^l e^{i  n\pi/2} \Phi (x^\mu, \LL, y) ~,
\nonumber  \\ [.3em]
&& \left.\partial_5\Phi \vb_{(x^4, x^5)=(y,0)}  = 
- e^{i  n\pi/2} 
\left.\partial_4\Phi \vb_{(x^4, x^5)=(0,y)} ~,
\nonumber  \\ [.3em]
&& \left.\partial_5\Phi \vb_{(x^4, x^5)=(y,\LL)}  = 
- (-1)^l e^{i  n\pi/2} 
\left.\partial_4\Phi \vb_{(x^4, x^5)=(\LL,y)} ~,
\label{summary-bc}
\eear
where $l,n$ are integers that can be restricted to $n= 0,1,2,3$ and
$l=0,1$.  It is interesting that the folding boundary conditions do
not depend on a continuous parameter, but rather there are only eight
self-consistent choices. 
Eqs.~(\ref{summary-bc}) include as particular cases the 
boundary conditions with vanishing $\Phi\partial_5 \Phi$ at $(y,0)$ or $(y,L)$.

Although two coefficients, $C_{1,2}^+$, remain unknown in the solution
(\ref{intermediate-solution}), they multiply two functions of $x^4$
and $x^5$ that either differ only by an interchange of $j$ and $k$ and
a factor of $e^{-i\theta}$, or are identical when $k=0$.  It turns out
that we can keep only one of these two coefficients, and still form a
complete set of functions on the square, which is the necessary and
sufficient condition for having a general Fourier decomposition as in
Eq.~(\ref{fourier}).  Furthermore, the normalization condition,
\be
\label{ortho}
\frac{1}{L^2}
\int_{0}^{\LL} dx^4 \int_{0}^{\LL} dx^5
\left[ f^{(j,k)}(x^4, x^5) \right]^* f^{(j^\prime,k^\prime)}(x^4, x^5)
= \delta_{j,j^\prime}\, \delta_{k,k^\prime}~,
\ee
determines the last coefficient up to a phase factor which we choose
to be one.  Explicitly, the solutions to Eq.~(\ref{twoD}) can be
written as
\bear
\label{KKsolns}
f_{n}^{(j,k)}(x^4, x^5) = \frac{1}{1+\delta_{j,0}\delta_{k,0} } 
\left[ e^{- i n \pi/2} \cos \pi\left( \frac{j x^4 + k x^5}{\LL} 
+ \frac{n}{2} \right)
+ \cos \pi\left( \frac{k x^4 - j x^5}{\LL} + \frac{n}{2} \right) 
\right] ~,
\nonumber \\ [.6em] {} \\ [-2em] {} \nonumber 
\eear
with $j + l/2$ and $k+l/2$ integers.

\begin{table}[t]
\centering
\renewcommand{\arraystretch}{1.5}
\begin{tabular}{|c| |c|c|c|c|c|c|c|c|c|}\hline
$(j,k)$ & (1,0) & (1,1) & (2,0) & $\ba{c} (2,1) \\(1,2) \ea $ & (2,2) 
& (3,0) & 
$\ba{c} (3,1) \\ (1,3) \ea $ & $\ba{c}  (3,2)\\ (2,3) \ea $ & (4,0)  
\\ \hline
$M_{j,k}R$
& 1 & $\sqrt{2}$ & 2 & $\sqrt{5}$ & $2\sqrt{2}$ & 3  &  
$\sqrt{10}$ & $\sqrt{13}$ & 4 
\\ \hline\hline\hline
$(j,k)$ 
& $\ba{c}  (4,1) \\ (1,4) \ea $& (3,3) & $\ba{c}  (4,2)\\ (2,4) \ea $ 
& 
$\ba{c}  (4,3)\\ (3,4) \\ (5,0) \ea $ & $\ba{c}  (5,1)\\ (1,5) \ea $ 
& 
$\ba{c}  (5,2)\\ (2,5) \ea $ & (4,4) & $\ba{c}  (5,3)\\ (3,5) \ea $ & 
(6,0)
\\ \hline
$M_{j,k}R$  & $\sqrt{17}$ & $3\sqrt{2}$ & $2\sqrt{5}$ & 5  
& $\sqrt{26}$ & $\sqrt{29}$  & 4 $\sqrt{2}$ & $\sqrt{34}$ & 6
\\ \hline
\end{tabular}
\bigskip
\caption{KK modes with $M_{j,k} \leq 6 M_{1,0}$, obtained for folding 
boundary
conditions with $l=0$. }
\label{TableKK1}  
\end{table}

The functions $f_{n}^{(j,k)}$ form a complete orthonormal set on the
square if
\be
\frac{1}{L^2}
\sum_{j,k} \left[ f_n^{(j,k)}(x^4, x^5) \right]^* 
f_n^{(j,k)}(x^{\prime 4}, x^{\prime 5}) 
= \delta(x^{\prime 4} - x^4) \, \delta(x^{\prime 5} - x^5) ~.
\label{completeness}
\ee
The allowed values for $j$ and $k$ must be chosen such that the above
completeness condition is satisfied.  In practice it is easier to
observe first that
\be
f_n^{(-j,k)} = (-1)^n f_n^{(j,-k)} = e^{i n \pi/2} f_n^{(k,j)} 
~,
\ee
so that it is sufficient to take $j > 0$, $k \geq 0$ and $j=k=0$.

Thus, for the  boundary conditions with $l = 0$, $j$ and $k$ 
take all integer values with
\be
j \geq 1 - \delta_{n,0}\delta_{k,0} \; , \;\; k \geq 0 ~.
\ee
One can check that the completeness condition
Eq.~(\ref{completeness}) is then satisfied.  We have obtained a tower
of four-dimensional fields $\Phi^{(j,k)}$ labeled by two integers,
with masses
\be
M^{(j,k)} = \sqrt{M_0^2 + \frac{j^2 + k^2}{R^2} } ~.
\ee
As usual, we will refer to these fields as KK modes.  The KK numbers,
$j$ and $k$, of the lightest KK modes are shown in Table 1.  For
$n=0$, there is an additional state: $j=k=0$.  This is a state of zero
momentum (``zero mode'') along both compact dimensions.

For the boundary conditions with $l = 1$, $j$ and $k$ take all
half-integer values satisfying
\be
j \geq \frac{1}{2} \; , \;\; k \geq \frac{1}{2} ~.
\ee 
In Table 2 are listed the KK numbers and masses of the lightest KK
modes in this case.

\begin{table}[t]
\centering
\renewcommand{\arraystretch}{1.5}
\begin{tabular}{|c| |c|c|c|c|c|c|c|}\hline
$(j,k)$ & {\large
$(\frac{1}{ 2},\frac{1}{2})$ } 
& {\large $\ba{c} (\frac{3}{2},\frac{1}{2}) \\ 
(\frac{1}{2},\frac{3}{2}) \ea $}
& {\large $(\frac{3}{2},\frac{3}{2})$ }
& {\large $\ba{c} (\frac{5}{2},\frac{1}{2}) \\ 
(\frac{1}{2},\frac{5}{2}) \ea $}
& {\large $ \ba{c} (\frac{5}{2},\frac{3}{2}) \\ 
(\frac{3}{2},\frac{5}{2}) \ea $}
& {\large $\ba{c} (\frac{5}{2},\frac{5}{2}) \\ 
(\frac{7}{2},\frac{1}{2}) 
    \\ (\frac{1}{2},\frac{7}{2}) \ea $ }
& {\large $\ba{c} (\frac{7}{2},\frac{3}{2}) \\ 
(\frac{3}{2},\frac{7}{2}) \ea $}
\\ \hline
$\sqrt{2} M_{j,k}R$
& $1$ & $\sqrt{5}$ & $3$ & $\sqrt{13}$ & $\sqrt{17}$ & $5$ & 
$\sqrt{29}$
\\ \hline
\end{tabular}
\bigskip
\caption{KK modes with $M_{j,k} \leq 6 M_{\frac{1}{2},\frac{1}{2}}$, 
obtained for folding boundary
conditions with $l=1$. }
\label{TableKK2} 
\end{table}

It is worth mentioning that field configurations may exist even when the 
boundary conditions (\ref{summary-bc}) are further restricted.
For example, the field derivatives vanish everywhere on the boundary 
provided $j$ and $k$ are integers and the KK functions are given by
either the orthonormal set consisting of $f_0^{(j,j)}$ and
\be
\label{coscos}
\frac{1}{\sqrt{2}}\left( f_0^{(j,k)} + f_0^{(k,j)} \right) 
= \sqrt{2} \left[ \cos \left(\frac{j x^4}{R} \right) \cos \left(\frac{k x^5}{R} \right) 
+ \cos\left(\frac{j x^4}{R} \right) \cos\left(\frac{k x^5}{R} \right) \right] 
\ee
with $j\neq k$, or 
\be
\frac{1}{\sqrt{2}}\left( f_2^{(j,k)} - f_2^{(k,j)} \right) 
= \sqrt{2}\left[\cos \left(\frac{j x^4}{R} \right) \cos \left(\frac{k x^5}{R} \right) 
- \cos\left(\frac{k x^4}{R} \right) \cos\left(\frac{j x^5}{R} \right)\right]   ~.
\ee
Note that the first of these KK towers has a zero mode,
namely $f_0^{(0,0)}$.
Another example is that where the field vanishes 
everywhere on the boundary, which requires $j$ and $k$ integers, 
and KK functions given by either 
\be
\frac{1}{\sqrt{2}}\left( f_0^{(j,k)} - f_0^{(k,j)} \right) 
= \sqrt{2}\left[ - \sin \left(\frac{j x^4}{R} \right) \sin \left(\frac{k x^5}{R} \right) 
+ \sin\left(\frac{k x^4}{R} \right) \sin\left(\frac{j x^5}{R} \right)\right]  ~,
\ee
or the orthonormal set consisting of $f_2^{(j,j)}$ and
\be
\label{sinsin}
-\frac{1}{\sqrt{2}}\left( f_2^{(j,k)} + f_2^{(k,j)} \right) 
= \sqrt{2}\left[\sin \left(\frac{j x^4 }{R}\right) \sin \left(\frac{k x^5}{R} \right) 
+ \sin \left(\frac{k x^4}{R} \right) \sin\left(\frac{j x^5}{R} \right)  \right]
\ee
with $j\neq k$.
Boundary conditions with the field derivatives vanishing on two 
sides of the square and the  field itself vanishing on the other two 
sides lead to the same  KK functions as in Eqs.~(\ref{coscos})-(\ref{sinsin})
but with $j$ and $k$ half-integers.

All the results obtained in this section for 
a complex scalar with $n=0$ or $n=2$
apply to the case of a real scalar field
as well (note that $f_{n}^{(j,k)}$ are complex functions for 
$n=1,3$).

\section{Fermions on a square: chiral boundary conditions}
\setcounter{equation}{0}

We now turn to free spin-1/2 fields in six dimensions.  The Clifford
algebra is generated by six anti-commuting matrices: $\Gamma^\alpha$,
$\alpha = 0,1, ...  5$.  The minimal dimensionality of these matrices
is $8\times 8$.  The $\Gamma$ matrices can be used to construct a
spinor representation of the $SO(1,5)$ Lorentz symmetry, with the
generators explicitly given by
\be
\frac{\Sigma^{\alpha\beta}}{2} = \frac{i}{4}[\Gamma^\alpha, 
\Gamma^\beta] ~.
\ee
This Lorentz representation is reducible and contains two irreducible
Weyl representations, which have different eigenvalues of the
chirality operator.  The two six-dimensional chiralities, labeled by
$+$ and $-$, are projected by the operators
\be
P_\pm = \frac{1}{2} \left( 1 \pm \bar{\Gamma} \right) ~,
\ee
where the six-dimensional chirality operator
\bear
\bar{\Gamma} & = & \frac{1}{6!} \epsilon_{\alpha_0\alpha_1 ... 
\alpha_5}
\Gamma^{\alpha_0}\Gamma^{\alpha_1} ... \Gamma^{\alpha_5} 
\nonumber \\ [0.2em]
& = & \Gamma^0\Gamma^1\Gamma^2\Gamma^3\Gamma^4\Gamma^5
\eear
is a self-adjoint matrix that anticommutes with all $\Gamma^\alpha$'s.
The chiral fermions in six dimensions have four components.

Upon compactification in the $x^4, x^5$ plane, the $SO(1,3)$ Lorentz
symmetry generated by $\Sigma^{\mu\nu}/2$, $\mu, \nu = 0,1,2,3$,
remains unbroken.
There are two chiralities under $SO(1,3)$, labeled as usual by 
$L$ and $R$. These are projected by 
\be
P_{L,R} = \frac{1}{2} \left( 1 \mp i\Gamma^0\Gamma^1\Gamma^2\Gamma^3 
\right)~.
\ee
A six-dimensional chiral fermion, $\Psi_\pm \equiv P_\pm \Psi$,
decomposes into two fermions of definite chirality under $SO(1,3)$:
\be
\Psi_\pm(x^\mu, x^4, x^5) = \Psi_{\pm_L}(x^\mu, x^4, x^5) + 
\Psi_{\pm_R}(x^\mu, x^4, x^5) ~,
\ee
where 
\be
\Psi_{\pm_{L,R}} \equiv P_{L,R} P_\pm \Psi  ~.
\ee
As in Section 2, we consider the compactification on a square: $0 \leq
x^4, x^5 \leq \LL$.  For definiteness, we analyze the case of a
 chirality  $+$ fermion.  At the end of this section we briefly comment
on the differences for the $-$ chirality.  The action for a free
six-dimensional chiral fermion is
\be
S_\Psi = \int d^4 x \int_0^{\LL} dx^4 \int_0^{\LL} dx^5
\frac{i}{2}
\left[\overline{\Psi}_+ \Gamma^\alpha \partial_\alpha \Psi_+ 
- \left(\partial_\alpha\overline{\Psi}_+\right) 
\Gamma^\alpha \Psi_+ \right] ~.
\label{fermion-action}
\ee
Under an arbitrary variation of the field, $\delta\Psi_+(x^\mu, x^4,
x^5)$, the action has to be stationary both inside the square and on
its boundary:
\bear
\delta S_\Psi^v & = & -  \int d^4 x 
\int_0^{\LL} dx^4 \int_0^{\LL} dx^5
i\left( \partial_\alpha \overline{\Psi}_+\right) 
\Gamma^\alpha \delta\Psi_+ = 0 ~,
\nonumber \\ [.5em]
\delta S_\Psi^s & = &  \frac{i}{2}\int d^4 x \left[
\int_0^{\LL} dx^4 \left(
\left. \overline{\Psi}_+ \Gamma^5 \delta \Psi_+\vb_{x^5=\LL}
- \left.\overline{\Psi}_+ \Gamma^5 \delta \Psi_+\vb_{x^5=0} \right) 
\right. \nonumber \\ [.3em] & & 
+ \left.\int_0^{\LL} dx^5  \left(
\left.\overline{\Psi}_+ \Gamma^4 \delta \Psi_+\vb_{x^4=\LL}
- \left.
\overline{\Psi}_+ \Gamma^4 \delta \Psi_+\vb_{x^4=0} \right) 
\right]  = 0 ~.
\label{fermion-vs}
\eear
The first equation implies that $\Psi_+$ is a solution to the
six-dimensional Weyl equation, which can be decomposed into two
equations:
\bear
 & & \Gamma^\mu \partial_\mu \Psi_{+_L} = - \left(\Gamma^4 \partial_4 
+ \Gamma^5 \partial_5 \right) \Psi_{+_R} ~,
\nonumber \\ [.3em] & & 
\Gamma^\mu \partial_\mu \Psi_{+_R} = - \left(\Gamma^4 \partial_4 
+ \Gamma^5 \partial_5 \right) \Psi_{+_L} ~.
\label{coupled-eqs}
\eear
The second equation (\ref{fermion-vs}) restricts the values of
$\Psi_+$ on the boundary:
\bear
&& \hspace*{-1.8em} 
\left.\overline{\Psi}_+ \Gamma^4 \,\delta \Psi_+ \vb_{(x^4,x^5)=(\LL, 
y)}
+ \left.\overline{\Psi}_+ \Gamma^5 \,\delta \Psi_+ \vb_{(x^4,x^5)=(y, 
\LL)}
\nonumber \\ [.4em] & & 
- \left.\overline{\Psi}_+ \Gamma^4  \, \delta \Psi_+ 
\vb_{(x^4,x^5)=(0,y)}
- \left.\overline{\Psi}_+ \Gamma^5 \, \delta \Psi_+ 
\vb_{(x^4,x^5)=(y,0)}
= 0 ~, 
\label{fermionBC}
\eear
for any $y \in [0,\LL]$.

We consider ``folding'' boundary conditions analogous to those imposed
on the scalar field in Section 2.1.  The important feature of these
boundary conditions is that they distinguish the four-dimensional
chiralities.  Explicitly, the phases that relate the fields on
adjacent sides of the square are different for left- and right-handed
fermions:
\bear
\Psi_{+_{L,R}}(x^\mu, y, 0) & = & e^{i \theta_{L,R}}
\Psi_{+_{L,R}} (x^\mu, 0, y) ~,
\nonumber \\ [.3em]
\Psi_{+_{L,R}}(x^\mu, y, \LL) & = & e^{i \tilde{\theta}_{L,R}}
\Psi_{+_{L,R}} (x^\mu, \LL, y)  ~,
\label{fermion-phase}
\eear
for any $y \in [0,\LL]$.  As in the case of scalars, the above
equations imply that
\bear
\left.\partial_4
\Psi_{+_{L,R}} \vb_{(x^4, x^5)=(y,0)} & = & e^{i \theta_{L,R}}
\left.\partial_5
\Psi_{+_{L,R}} \vb_{(x^4, x^5)=(0,y)} ~,
\nonumber \\ [.3em] 
\left.\partial_4
\Psi_{+_{L,R}} \vb_{(x^4, x^5)=(y, \LL)} & = & e^{i 
\tilde{\theta}_{L,R}} 
\left.\partial_5
\Psi_{+_{L,R}} \vb_{(x^4, x^5)=(\LL, y)} ~.
\label{fermion-auto}
\eear

The study of complex scalars presented in Section 2 has shown that
additional boundary conditions, Eqs.~(\ref{der-phase}), must be
imposed on the field derivatives in order to have the same Lagrangians
at the identified points.  We now show that in the case of fermions
the boundary conditions Eqs.~(\ref{fermion-phase}) are sufficient to
ensure the equality of the Lagrangians at the identified points.  To
this end, we write the first equation of motion (\ref{coupled-eqs}) at
$(x^4,x^5) = (0,y)$, multiply it by $e^{i\theta_L}$, and subtract the
same equation evaluated at $(x^4,x^5) = (y,0)$.  Using then
Eqs.~(\ref{fermion-auto}) and the identities
\bear
\label{Pidentities}
\Gamma^4 P_L P_\pm  = \mp i \Gamma^5 P_L P_\pm 
 ~,
\nonumber \\ [.3em] 
\Gamma^4 P_R P_\pm  = \pm i \Gamma^5 P_R P_\pm  ~,
\eear
we obtain
\bear
\label{first-eom}
\left.\partial_4 \Psi_{+_R} \vb_{(x^4, x^5)=(0, y)}
+ i e^{- i\theta_L}
\left.\partial_5 \Psi_{+_R} \vb_{(x^4, x^5)=(y,0)}
= i \left[ 1 - i e^{i(\theta_R - \theta_L)} \right]
\left.\partial_5 \Psi_{+_R} \vb_{(x^4, x^5)=(0, y)} ~.
\nonumber \\ [.6em] {} \\ [-2em] {} \nonumber 
\eear
Writing the second equation of motion (\ref{coupled-eqs}) at 
$(x^4,x^5)
= (0,y)$, multiplying it by $e^{i\theta_R}$, and subtracting the same
equation evaluated at $(x^4,x^5) = (y,0)$ gives
\bear
\label{second-eom}
\left.\partial_4 \Psi_{+_L} \vb_{(x^4, x^5)=(0, y)}
- i e^{- i\theta_R}
\left.\partial_5 \Psi_{+_L} \vb_{(x^4, x^5)=(y,0)}
= - i \left[ 1 + i e^{-i(\theta_R - \theta_L)} \right]
\left.\partial_5 \Psi_{+_L} \vb_{(x^4, x^5)=(0, y)} ~.
\nonumber \\ [.6em] {} \\ [-2em] {} \nonumber 
\eear
Based on the above two equations, one can check that the Lagrangian at
$(x^4,x^5) = (y,0)$ is equal to the one at $(x^4,x^5) = (0,y)$.

The same procedure gives two equations for the derivatives at
$(x^4,x^5) = (L,y)$ and $(y,L)$ which are analogous to
Eqs.~(\ref{first-eom}) and (\ref{second-eom}), except for the
replacement of $\theta_{L,R}$ by $\tilde{\theta}_{L,R}$.  As a result,
the Lagrangians at $(x^4,x^5) = (L,y)$ and $(y,L)$ are also equal.

The boundary conditions (\ref{fermion-phase}) must satisfy the
stationarity condition (\ref{fermionBC}).  Given that the field
variations on opposite sides of the square are independent, and that
the variations of $\Psi_{+_L}$ and $\Psi_{+_R}$ are independent at any
point, Eqs.~(\ref{fermionBC}) are satisfied if and only if
\bear
&& 
\left[ 1 - i e^{i (\theta_R - \theta_L)} \right]
\overline{\Psi}_{+_L} (x^\mu, 0, y) \Gamma^4 \Psi_{+_R}(x^\mu, 0, y)
= 0~,
\nonumber \\ [.3em] 
&& \left[ 1 - i e^{i (\tilde{\theta}_R - \tilde{\theta}_L)} \right]
\overline{\Psi}_{+_L} (x^\mu, \LL, y) \Gamma^4               
\Psi_{+_R} (x^\mu,\LL, y)
=0~.
\label{stationary-final}
\eear

We now show that Eqs.~(\ref{fermion-phase}), (\ref{first-eom}),
(\ref{second-eom}) and (\ref{stationary-final}) imply that both
$\Psi_{+_L}$ and $\Psi_{+_R}$ satisfy the same boundary conditions as
in the scalar case of Section 2.  To see this, note first that for
\be
e^{i(\theta_L - \theta_R )} = i 
\label{LR-phases}
\ee
Eqs.~(\ref{first-eom}) and (\ref{second-eom}) take the form
\be
\left.\partial_4 \Psi_{+_{L,R}} \vb_{(x^4, x^5)=(0, y)}
= - e^{- i\theta_{L,R}}
\left.\partial_5 \Psi_{+_{L,R}} \vb_{(x^4, x^5)=(y,0)} ~,
\label{fermion-phase2}
\ee
which is the same as the boundary condition for the scalar
derivatives, Eq.~(\ref{der-phase}), with the identification $e^{i
\theta'} = - e^{i \theta_{L,R}}$.  Therefore, in this case we
automatically obtain the restriction given by the relation
(\ref{phase-relation}) between $\theta$ and $\theta^\prime$.

If Eq.~(\ref{LR-phases}) is not assumed, then
Eq.~(\ref{stationary-final}) requires that either
$\Psi_{+_R}$ or $\Psi_{+_L}$ vanish at the $(0, y)$ points.
These two cases yield the same result, so we
present here only the $\Psi_{+_R}(0, y) = 0$ case.
Then, $\partial_5\Psi_{+_R}|_{(x4, x5)=(0,y)}=0$ so that
Eq.~(\ref{first-eom}) implies that the boundary condition
(\ref{fermion-phase2}) for $\Psi_{+_R}$ is satisfied,
with $\theta_R$ defined in terms of $\theta_L$ by Eq.~(\ref{LR-phases}).
Furthermore, $\Psi_{+_R}(y, 0) = 0$ due to the
folding boundary condition (\ref{fermion-phase}).
The six-dimensional Dirac equation (\ref{coupled-eqs}), together with the
identity (\ref{Pidentities}), then implies that
\be
\label{derivativeconditions}
(\partial_4 + i \partial_5)\Psi_{+_L} |_{(x4, x5)=(y,0)} =
(\partial_4 + i \partial_5)\Psi_{+_L} |_{(x4, x5)=(0,y)} = 0~.
\ee
These two equations together with the folding boundary condition for
$\Psi_{+_L}$ lead again to Eq.~(\ref{fermion-phase2}) for
$\Psi_{+_L}$.

Similar conclusions
(with $\theta_{L,R}$ replaced by $\tilde{\theta}_{L,R}$) apply 
at $(L, y)$ and $(y, L)$:
\be
e^{i(\tilde{\theta}_L - \tilde{\theta}_R )} = i ~,
\label{LR-phases-other}
\ee
and the boundary conditions are those encountered in the scalar case.

We can now solve the six-dimensional Weyl equation.  First we observe
that the coupled first-order equations (\ref{coupled-eqs}) imply that
both $\Psi_{+_L}$ and $\Psi_{+_R}$ obey the six-dimensional
Klein-Gordon equation for a massless field.  Therefore, similarly to
the case of a scalar discussed in Section 2, $\Psi_{+_L}$ and
$\Psi_{+_R}$ can be decomposed in Fourier modes.  However, further
complications arise because the Weyl fermions in six dimensions have
four components, while in four dimensions they have two components.
The $\Gamma$ matrices can be written as the direct product of the
usual four-dimensional $\gamma$ matrices with the Pauli matrices:
\be
\Gamma^\mu = \gamma^\mu \otimes \sigma^0 \; \; , \; \;
\Gamma^{4,5} = i \gamma_5 \otimes \sigma^{1,2} ~,
\ee
where $\sigma^0$ is the $2\times 2$ unit matrix, and 
$\gamma_5 = i\gamma^0\gamma^1\gamma^2\gamma^3$.  Note that
$\overline{\Gamma} = -\gamma_5 \otimes \sigma^3$.  The KK
decomposition can then be written as follows:
\bear
\Psi_{+_L}(x^\mu, x^4, x^5) = \frac{1}{L}
\sum_{j,k}  f_{+_{L}}^{(j,k)}(x^4, x^5) \Psi_{+_L}^{(j,k)} (x^\mu)
 \otimes \left(\ba{c} 1 \\ 0 \ea \right)~,
\nonumber \\ [1em]
\Psi_{+_R}(x^\mu, x^4, x^5) = \frac{1}{L}
\sum_{j,k}  f_{+_{R}}^{(j,k)}(x^4, x^5) \Psi_{+_R}^{(j,k)} (x^\mu)
 \otimes \left(\ba{c} 0 \\ 1 \ea \right)~,
\label{fermion-fourier}
\label{fermionexpansion}
\eear
where $f_{+_{L,R}}^{(j,k)}$ are scalar functions of $x^4$ and $x^5$, while
$\Psi_{+_{L,R}}^{(j,k)}$ are Weyl fermions in four dimensions.  The
$\gamma$ matrices act on $\Psi_{+_{L,R}}^{(j,k)}$, and the Pauli
matrices act on the 2-columns shown in the above equation.  The
$\Psi_{+_L}^{(j,k)}$ and $\Psi_{+_R}^{(j,k)}$ fields form a Dirac
fermion in four dimensions
\be
\left( i \gamma^\mu \partial_\mu  - M_{j,k}\right) 
\left( \Psi_{+_L}^{(j,k)} + \Psi_{+_R}^{(j,k)} \right) = 0 ~,
\ee
which explains why the real numbers $j$ and $k$ are the same for the
KK decompositions of both $\Psi_{+_L}^{(j,k)}$ and
$\Psi_{+_R}^{(j,k)}$ shown in Eq.~(\ref{fermion-fourier}).  The mass
$M_{j,k}$ turns out to be the same as in the scalar case, {\it i.e.},
it is given by Eq.~(\ref{mjk}).

Inserting the KK decomposition in the first-order equations
(\ref{coupled-eqs}) we obtain that $f_{+_{L,R}}^{(j,k)}$ must be
solutions to
\bear 
&& \left(\partial_4 - i \partial_5\right) f_{+_{R}}^{(j,k)}
= M_{j,k}f_{+_{L}}^{(j,k)} ~,
\nonumber \\ [1em]
&& \left(\partial_4 + i \partial_5\right) f_{+_{L}}^{(j,k)}
= -M_{j,k}f_{+_{R}}^{(j,k)} ~.
\label{f-coupled}
\eear
Acting on the first equation with $\partial_4 + i \partial_5$ and then
using the second equation we find that $f_{+_{R}}$ satisfies the
second-order equation encountered in the scalar case,
Eq.~(\ref{twoD}).  Having also the same boundary conditions as for the
scalars, as we discussed above, implies that $f_{+_{R}}^{(j,k)}$ is
given by the right-hand side of Eq.~(\ref{KKsolns}).  Therefore,
$f_{+_{R}}^{(j,k)}$ depends on an integer that can take four values,
$n^+_R = 0,1,2,3$, and for each of these cases there is a solution
where $j$ and $k$ are integers (a case labeled by $l^+ = 0$), and a
different solution ($l^+ = 1$) where they are half-integers.  Given
one of these solutions for $f_{+_{R}}^{(j,k)}$, the first equation in
(\ref{f-coupled}) determines $f_{+_{L}}^{(j,k)}$ and gives a solution
of the same form except for a shift by one in $n^+_R$ and an overall
phase factor:
\bear
\label{KKsolns-fermion}
&& f_{+_R}^{(j,k)} = f_{n^+_R}^{(j,k)} ~,
\nonumber \\ [1em]
&& f_{+_L}^{(j,k)} = 
\frac{k+ij}{\sqrt{j^2+k^2}}\, f_{1+n^+_R}^{(j,k)} ~,
\eear 
where $f_n^{(j,k)}$ is given by Eq.~(\ref{KKsolns}).  

To summarize,
a chirality $+$ fermion with an expansion in terms of the above KK
wavefunctions, as in Eq.~(\ref{fermionexpansion}), satisfies the
boundary conditions
\bear
\label{summary-Rbc}
&& \Psi_{+_R}(x^\mu, y, 0) = e^{i n^+_R \pi/2} \Psi_{+_R}(x^\mu, 0, 
y) ~,
\nonumber \\ [.3em]
&& \Psi_{+_R}(x^\mu, y, \LL) = (-1)^{l^+} e^{i  n^+_R \pi/2} 
\Psi_{+_R}(x^\mu, \LL, y) ~,
\nonumber  \\ [.3em]
&& \left.\partial_5\Psi_{+_R} \vb_{(x^4, x^5)=(y,0)}  = 
- e^{i  n^+_R \pi/2} 
\left.\partial_4\Psi_{+_R} \vb_{(x^4, x^5)=(0,y)} ~,
\nonumber  \\ [.3em]
&& \left.\partial_5\Psi_{+_R} \vb_{(x^4, x^5)=(y,\LL)} = 
- (-1)^{l^+} e^{i  n^+_R \pi/2} 
\left.\partial_4\Psi_{+_R} \vb_{(x^4, x^5)=(\LL,y)} ~,
\eear
for the right-handed component, while for the 
left-handed component the same boundary conditions
apply, except for $n^+_R$ being replaced by 
\be
n^+_L = n^+_R + 1 \; {\rm mod} \, 4~.
\ee

The $-$ chirality can be treated in an analogous fashion.  The
only difference compared to the $+$ chirality  discussed above arises from
Eq.~(\ref{Pidentities}), and leads to the interchange of $L$ and $R$ in
the equations following Eq.~(\ref{Pidentities}).
Thus, $\Psi_{-_L}$ and $\Psi_{-_R}$
have folding boundary conditions characterized by two integers each, 
$n^{-}_L, l^{-}$ and $n^{-}_R, l^{-}$, respectively, with 
$n^{-}_L, n^{-}_R = 0,1,2,3$, $l^{-}=0,1$, and 
\be
n^-_L = n^-_R - 1 \; {\rm mod} \, 4~.
\ee
Their KK wave functions are related as follows:
\bear
\label{KKsolns-fermion-other}
&& f_{-_L}^{(j,k)} =  f_{n^-_L}^{(j,k)} ~,
\nonumber \\ [1em]
&& f_{-_R}^{(j,k)} = 
\frac{k+ij}{\sqrt{j^2+k^2}}\, f_{1+n^-_L}^{(j,k)} ~.
\eear 

The KK spectrum of a six-dimensional chiral fermion consists of four-dimensional
vector-like fermions of masses $M_{jk}$, as in Tables 1 and 2. 
Only the zero-modes are four-dimensional chiral fermions. 
It is interesting that there are two kinds of zero-modes for each
four-dimensional chirality: depending on whether the six-dimensional fermion 
has chirality  $+$ or $-$, the tower of KK modes that includes a left-handed
zero-mode is paired with a tower of KK modes with wavefunctions of the
$f_3$ or $f_1$ type, and vice-versa for a right-handed zero-mode.

The discussion in this section has been restricted so far to chiral
six-dimensional fermions. It is also useful to analyze the case of a vector-like 
six-dimensional fermion, $\Psi$, of mass $M_0$. 
The boundary conditions and solutions to the field equations 
for $\Psi_+$ and $\Psi_-$, obtained independently above, continue to apply
when both chiralities are present, provided the equality of 
Lagrangians at identified points is satisfied. 
Therefore, the presence in the Lagrangian of the mass term 
\be
-M_0 \overline{\Psi} \Psi
= -M_0 \left( \overline{\Psi}_{+_L} \Psi_{-_R} + 
\overline{\Psi}_{+_R} \Psi_{-_L} \right) + 
{\rm h.c.} 
\ee
relates the KK wave functions of the 
$\Psi_+$ and $\Psi_-$ components of $\Psi$.
The mass term is the same at identified boundary 
points when $n^\pm_L = n^\mp_R$.
Another implication of the six-dimensional mass term is that
the $M_{j,k}$ mass of a KK mode is now replaced by a matrix:
\be
\left( \overline{\Psi}^{(j,k)}_{+_L}\, , \, 
\overline{\Psi}^{(j,k)}_{-_L} \right)
\left(  \ba{cc}  M_{j,k} & M_0 \\ M_0^* & -M_{j,k} \ea \right)
\left( \ba{c} \Psi^{(j,k)}_{+_R} \\ [.5em]
\Psi^{(j,k)}_{-_R} \ea \right) ~.
\ee
Both eigenvalues are equal to $(M_0^2 + M_{j,k}^2)^{1/2}$,
which is the same as the mass of the $(j,k)$ mode of a scalar with 
bulk mass $M_0$.
Thus, a vector-like six-dimensional fermion includes
two degenerate towers of vector-like KK modes, and at most a single 
vector-like $j=k=0$ state of mass $M_0$ (when $n^+_L n^-_L = 0$).
Note that this is also the case for the $T^2/Z_2$ orbifold 
\cite{Appelquist:2000nn},
and that the degeneracy is lifted by loops if 
$\Psi_+$ and $\Psi_-$ have different interactions \cite{Cheng:2002iz}.

\section{Symmetries}
\setcounter{equation}{0}

So far we have analyzed free scalar and fermion fields, and found that
the equations of motion have nontrivial solutions only if the phases
associated with the folding boundary conditions are restricted to a
discrete set of values.  In this section we show that those
restrictions lead to the existence of certain symmetries that are
obeyed in theories with any number of scalar and fermion fields, and
with any type of local six-dimensional Lorentz-invariant interactions.

\subsection{$Z_8 \times \Ztwol$ invariance}

Consider a theory with a number of six-dimensional complex scalar
fields, $\Phi_i(x^\mu, x^4, x^5)$ with $i=1,..., p$.  By studying
the free part of the Lagrangian, as we did above, each of these is
subject to folding boundary conditions as in Eq.~(\ref{summary-bc}),
but the integers $n_{i} = 0,1,2,3$ and $l_{i}= 0,1$ that determine the
boundary conditions may differ for different fields, and are therefore
labeled by a flavor 
index $i$.\footnote{If the $\Phi_{i}$ span a 
representation of a non-abelian internal symmetry, 
it may be possible to impose 
Eqs.~(\ref{phase}) and (\ref{other-phase}) 
with $\Phi = (\Phi_{1}, \ldots, \Phi_{p})$ 
and the phase $e^{i\theta}$ replaced by a matrix. 
This may be interesting as a higher 
dimensional mechanism for symmetry breaking 
\cite{orbifoldbreaking}, but we do not 
consider this possibility here.}  
The most general interactions involving
these fields which do not involve derivatives are of the type
\be
\label{nonderivativeoperators}
\prod_{i = 1}^{p} (\Phi_i)^{m_i}  (\Phi_i^\dagger)^{m_i^\prime} ~,
\ee
where $m_i, m_i^\prime \ge 0$ are integers.  The key fact, which
follows from the form of the folding boundary conditions,
Eq.~(\ref{summary-bc}), is that equality of the Lagrangians at the
identified boundary points $(y,0)$ and $(0,y)$ requires that the
overall phase difference is a multiple of $2\pi$:
\be
\label{nselection}
\sum_{i = 1}^{p} n_i (m_i - m_i^\prime)  = 0 \; {\rm mod } \, 4 ~.
\ee
This equation implies that all such interaction terms are invariant
under the $Z_4$ transformations
\be
\label{Z4symmetry}
\Phi_i(x^\mu, x^4, x^5) \mapsto e^{-i n_i \pi/2} \Phi_i(x^\mu, x^4, 
x^5) ~.
\ee
Furthermore, equality of the Lagrangians at the identified boundary
points $(y,L)$ and $(L,y)$ requires in addition
\be
\label{lselection}
\sum_{i = 1}^{p} l_i (m_i - m_i^\prime)  = 0 \; {\rm mod } \, 2 ~,
\ee
so that the operators (\ref{nonderivativeoperators}) are 
invariant under the additional $Z_{2}$ transformations
\be
\label{Z2symmetry}
\Phi_i(x^\mu, x^4, x^5) \mapsto (-1)^{l_{i}} \Phi_i(x^\mu, x^4, 
x^5) ~.
\ee
We will refer to this $Z_{2}$ symmetry as $\Ztwol$.

The six-dimensional Lorentz invariance ensures that  the operators
that include derivatives in the most general way are also invariant under
the above $Z_{4}$ and $\Ztwol$ transformations.
To see this, recall first that only
the derivatives of a field along the compact dimensions have boundary
conditions with an $n$ integer different than for the field itself:
\be
\left. \left(\partial_4 \pm i \partial_5 \right) \Phi_i 
\right|_{(x^4,x^5)=(y,0)} =
e^{i (n_{i} \mp 1)\pi/2}
\left. \left(\partial_4 \pm i \partial_5 \right) \Phi_i 
\right|_{(x^4,x^5)=(0,y)} ~.
\label{pm-der}
\ee
Six-dimensional Lorentz invariance allows only two types of
combinations of derivatives,
\bear
&&
\partial^\alpha \Phi_1\partial_\alpha \Phi_2 =
\partial^\mu \Phi_1\partial_\mu \Phi_2
+ \frac{1}{2}\left[ \left(\partial_4 + i \partial_5 \right) 
\Phi_1\left(\partial_4 - i \partial_5 \right) \Phi_2
+ \left(\partial_4 - i \partial_5 \right) 
\Phi_1\left(\partial_4 + i \partial_5 \right) \Phi_2 \right]
\nonumber \\ [.5em]
&&
\epsilon^{\alpha_1 ... \alpha_6}
\partial_{\alpha_1} \Phi_1...\partial_{\alpha_6} \Phi_6
= \frac{i}{2}\sum \epsilon^{\mu_1 ... \mu_4}
\partial_{\mu_1} \Phi_{i_1}...\partial_{\mu_4} \Phi_{i_4}
\nonumber \\ [.3em] 
&& \hspace*{6em}
\times \left[ \left(\partial_4 + i \partial_5 \right) \Phi_{i_5}
\left(\partial_4 - i \partial_5 \right) \Phi_{i_6}
- \left(\partial_4 - i \partial_5 \right) \Phi_{i_5}
\left(\partial_4 + i \partial_5 \right) \Phi_{i_6} \right] ~,
\eear
where the sum in the right-hand side of the second equation is over
the permutations of the set of indices $\{i_1, ...  ,i_6\} = \{1,...,
6\}$.  From Eq.~(\ref{pm-der}) then follows that even in the presence
of derivatives the equality of the Lagrangians at the identified
boundary points requires that Eqs.~(\ref{nselection}) and
(\ref{lselection}) be satisfied.  Therefore, the scalar action is
invariant under a $Z_4\times \Ztwol$ symmetry defined by
Eqs.~(\ref{Z4symmetry}) and (\ref{Z2symmetry}).  We will refer to the
$n_i$ and $l_i$ that characterize the folding boundary conditions of
$\Phi_i$ as the $Z_4\times \Ztwol$ charges of $\Phi_i$.

In the case of fermions, the same results clearly apply to operators
where all the derivatives have Lorentz indices contracted among
themselves.  However, the gamma matrices also carry Lorentz indices,
and when these are contracted with the indices of the derivatives the
equality of the Lagrangians at $(y,0)$ and $(0,y)$ leads to additional
constraints.  For example, a kinetic term $i\overline{\Psi}_{+}^i
\Gamma^\alpha \partial_\alpha \Psi_{+}^i$, where the upper index $i$ 
labels different flavors, includes a piece
\be
i\overline{\Psi}_{+_L}^i \left(\Gamma^4 + i \Gamma^5\right)
\left(\partial_4 - i \partial_5 \right) \Psi_{+_R}^i
\ee
which is consistent with the folding boundary conditions only if
$n_{Ri}^+ - n_{Li}^+ = -1$ [see Eq.~(\ref{summary-Rbc})].  Thus, for
fermions, the naive $Z_4$ transformation Eq.~(\ref{Z4symmetry}) is not
a symmetry.  

However, the fermion kinetic term is invariant under the
$Z_{8}$ transformations
\bear
\label{componentZ8symmetry}
\Psi^{i}_{\pm R}(x^\mu, x^4, x^5) &\mapsto& e^{-i (\pm 1/2+n^{\pm}_{Ri})
\pi/2} \Psi^{i}_{\pm R}(x^\mu, x^4, x^5) ~,
\nonumber \\
\Psi^{i}_{\pm L}(x^\mu, x^4, x^5) &\mapsto& e^{-i (\mp 1/2+n^{\pm}_{Li})
\pi/2} \Psi^{i}_{\pm L}(x^\mu, x^4, x^5) ~,
\eear
where we also included the transformations for the $-$ chirality.  To
see that the above transformation is a symmetry of a theory 
with any interactions,
we first note that it can be written as
\bear
\label{Z8symmetry}
\Psi^i(x^\mu, x^4, x^5) &\mapsto& {\Psi'}^i(x^\mu, x^4, x^5)
\equiv e^{-i (\pi/2) \Sigma_{45}/2} \Psi^i(x^\mu, -x^5, x^4) ~,
\eear
where $\Psi^i$ is a six-dimensional spinor and we used the fact
that the fermion fields on the folded square satisfy
\bear
\label{rotationproperty}
\Psi^{i}_{\pm R}(x^\mu, -x^5, x^4) &=& e^{-i n^{\pm}_{ Ri} \pi/2}
\Psi^{i}_{\pm R}(x^\mu, x^4, x^5) ~,
\nonumber \\
\Psi^{i}_{\pm L}(x^\mu, -x^5, x^4) &=& e^{-i n^{\pm}_{Li} \pi/2}
\Psi^{i}_{\pm L}(x^\mu, x^4, x^5) ~,
\eear
which follow from the corresponding property of the KK wavefunctions
in Eq.~(\ref{KKsolns}), 
\be
f_n^{(j,k)}(-x^5, x^4) = e^{-i n \pi/2}
f_n^{(j,k)}(x^4, x^5) ~.  
\ee
It is now clear that the
transformations~(\ref{componentZ8symmetry}) correspond to a rotation
by $\pi/2$ in the plane of the compact dimensions around the point
$(x^4, x^5) = (0, 0)$, given that the fields initially defined
on the square $0<x^{4},x^{5} <L$ can be 
analytically continued to the whole plane.

All {\textit{local}} operators in the (compactified) six-dimensional
theory are restricted by the six-dimensional Lorentz symmetry and in
particular by the transformations (\ref{Z8symmetry}).  More precisely,
such operators satisfy
\be
\label{allowedoperators}
{\cal{O}}[\overline{\Psi'}^{i_1}(x^\alpha), {\Psi'}^{i_2}(x^\alpha),
\partial / \partial x^\alpha ] 
= {\cal{O}}[\overline{\Psi}^{i_1}(x^{\prime\alpha}),  {\Psi}^{i_2}(x^{\prime\alpha}),
\partial / \partial x^{\prime\alpha} ] 
\equiv {\cal{O}}(x')
\ee
where ${\Psi'}^{i}(x^{\alpha})$ are the Lorentz transformed fields.
Equation~(\ref{allowedoperators}) is the statement that the local
operators that may appear in the six-dimensional Lagrangian are
Lorentz scalars. The six-dimensional Lorentz symmetry is 
broken by compactification, but this does not relax 
the restriction (\ref{allowedoperators}) 
on local operators.  For the transformation
(\ref{Z8symmetry}), the left-hand side in Eq.~(\ref{allowedoperators})
can be written as
\be
\label{lhs}
{\cal{O}}[\overline{{\Psi'}}^{i_1}_{\pm R,L}(x^\alpha),
{\Psi'}^{i_2}_{\pm R,L}(x^\alpha),\partial / \partial x^\alpha] =
e^{i (\pi/2) \left(\sum_{i_2} q_{i_2} - \sum_{i_1} q_{i_1}\right)}
 {\cal{O}}(x) ~,
\ee
where $q_i = \pm 1/2 + n_i$ are the appropriate charges as defined in
Eq.~(\ref{componentZ8symmetry}).  In addition, the equality of the
Lagrangians at $(0,y)$ and $(y,0)$ implies that
\be
\left. {\cal{O}}(x')\right|_{(x^{\prime 4}, x^{\prime 5}) = (0,y)}
= \left. {\cal{O}}(x)\right|_{(x^4, x^5) = (y,0)} ~,
\ee
which together with Eqs.~(\ref{allowedoperators}) and (\ref{lhs})
requires
\be
\label{chargeconservation}
\sum_{i_2} q_{i_2} - \sum_{i_1} q_{i_1} = 0 \; {\rm mod } \, 4 ~.
\ee
Therefore, imposing the equality of the Lagrangians at $(y,0)$ and
$(0,y)$ relates the integers $n_{i}$ that characterize the boundary
conditions for the various fields in such a way that the theory is
invariant under (\ref{componentZ8symmetry}).

Finally, as in the scalar case, equality of the Lagrangians at the
$(y,\LL)$ and $(\LL,y)$ boundaries implies that the theory is
invariant under the $\Ztwol$ transformation
\be
\label{Z2symmetryfermions}
\Psi^{i}_{\pm}(x^\mu, x^4, x^5) \mapsto (-1)^{l^{\pm}} 
\Psi^{i}_{\pm}(x^\mu, x^4, x^5) ~,
\ee
where the same $l^{\pm}$ applies to both four-dimensional chiralities
$L$ and $R$ belonging to a given six-dimensional fermion $\Psi^i_{\pm}$.

In general, quantum loops  contain
divergences which correspond to localized counterterms at the points
$(0,0)$ and $(\LL,\LL)$.  This is similar to the situation of a fifth
dimension compactified on the interval (the $S^{1}/Z_{2}$ orbifold), where
localized operators are generated at the boundaries \cite{Georgi:2000ks}.  
Furthermore,
the six-dimensional theory could include such localized operators 
at tree level, 
with coefficients determined by the matching, at the
six-dimensional cutoff scale $\Lambda$, between the compactified
six-dimensional theory and its UV completion.  One may worry that such
operators could violate the symmetries discussed above.  We note,
however, that by virtue of their locality such operators are still
tightly constrained by the six-dimensional spatial symmetries.  What
is special about the ``fixed points'' $(0,0)$ and $(\LL,\LL)$ is that
they correspond to the location of physical objects, or ``branes'',
with attributes such as tension, etc.  In particular, their presence
allows to distinguish the four dimensions parallel to the brane from
the two dimensions transverse to it, breaking the
six-dimensional Lorentz invariance.\footnote{The branes, and the
operators localized on them, should be reparametrization invariant
under both six-dimensional coordinate, as well as brane
(``worldsheet'') coordinate transformations.  This is simply the
statement that the brane, as a physical object, should have a
geometric description.  Therefore, there is a sense in which these
operators are still constrained by local six-dimensional Lorentz
transformations.} The key point, however, is that in our
compactification the space in the vicinity of the ``branes'' has a
rotational symmetry in the transverse dimensions.  In fact, the brane
locations correspond to conical singularities, with deficit angles of
$3\pi/2$.  Thus, the brane-localized operators should be explicitly
$SO(3,1)\times SO(2)$ invariant.  Since our argument above,
Eqs.~(\ref{allowedoperators})--(\ref{chargeconservation}), was based only 
on the $SO(2)$ rotational symmetry, we conclude as before that its
$Z_{8}$ subgroup is an exact symmetry of the compactified theory.  The
$\Ztwol$ symmetry associated with fields with $l^{\pm} = 1$ is clearly
also a symmetry of the localized operators.
The other corners of the square, $(L,0)$ and $(0,L)$, which are 
identified, have a conical singularity with a deficit angle of
$\pi$, and the $SO(2)$ rotational symmetry also ensures that
any operators localized there are invariant under $Z_8$. 

We have shown that the fact that there are eight possible boundary
conditions that allow a nontrivial solution to the equation of motion
leads to a $Z_8 \times \Ztwol$ symmetry.  Under the $Z_8$ symmetry a
fermion of chirality $+_R$ or $-_L$ has charge $1/2 + n$, a fermion of
chirality $-_R$ or $+_L$ has charges $-1/2 + n$, and a scalar has
charge $n$, where $n=0,1,2,3$ defines the boundary conditions
(\ref{summary-bc}) on two adjacent sides of the square.  Any operator
is $Z_8$ invariant if the total charge is a multiple of four.  Under
the $\Ztwol$ symmetry, any field has a charge $l=0,1$, where $l$ fixes
the relative sign between the boundary conditions on the two pairs of
identified sides of the square, as in Eq.~(\ref{summary-bc}).

\subsection{Kaluza-Klein parity}

Next we will show that 
besides the $Z_8 \times \Ztwol$ symmetry discussed so far 
there is another discrete symmetry that restricts the interactions 
among KK modes.

Let us first concentrate on theories where all fields 
have $l = 0$ (but with no restriction on $n$). 
Under reflections (or, equivalently, rotations by $\pi$) about the 
center of the square $(\LL/2,\LL/2)$,
\be
\label{reflection}
(x^4, x^5) \mapsto (\LL - x^4, \LL - x^5)~,
\ee
the folding boundary conditions [see Eqs.~(\ref{summary-bc}) 
and (\ref{summary-Rbc})] are invariant 
because the conditions that relate $l=0$ fields 
at $(0,y)$ and $(y,0)$ are interchanged
with the ones at $(\LL,y)$ and $(y,\LL)$.
Furthermore, the six-dimensional Lagrangian is invariant under
\bear
\Phi(x^\mu, x^{4},x^{5}) &\mapsto& \Phi(x^\mu, \LL - x^{4},\LL - 
x^{5}) \hspace{2.9cm} \textrm{(scalars)}~, \nonumber
\\
\Psi_{\pm}(x^\mu, x^{4},x^{5}) &\mapsto& e^{-i \pi \Sigma_{45}/2} \, 
\Psi_{\pm}(x^\mu, \LL - x^{4},\LL - x^{5}) \hspace{1cm} 
\textrm{(fermions)}~,  
\eear
where the phases are given  by $\mp i$
for $\Psi_{\pm R}$ and $\pm i$ for $ \Psi_{\pm L}$. 
Thus, the six-dimensional action is also invariant under
the above transformations.
 Consequently,
 the KK wavefunctions of type $n$ are mapped into
fields of the same type $n$. Explicitly,  Eq.~(\ref{KKsolns}) implies
\be
f_n^{(j,k)}(\LL - x^4, \LL - x^5) = (-1)^{j+k+n} f_n^{(j,k)}(x^4, 
x^5)~.
\ee
It follows that all bulk interactions, when decomposed
into KK modes, give rise to interaction terms invariant under
\bear
\Phi^{(j,k)} (x^\mu) &\mapsto& (-1)^{j+k+n} \, \Phi^{(j,k)} (x^\mu) ~,
\\
\Psi_{\pm R,L}^{(j,k)} (x^\mu) &\mapsto& 
(-1)^{j+k+q} \, \Psi_{\pm R,L}^{(j,k)} (x^\mu) ~,
\eear
where $\Phi^{(j,k)} (x^\mu)$, $\Psi_{\pm R,L}^{(j,k)} (x^\mu)$ are the
KK modes as defined in Eqs.~(\ref{fourier}) and
(\ref{fermionexpansion}), and the charges for the fermions, $q = \pm
1/2 +n$, are as given in Eq.~(\ref{componentZ8symmetry}).  Since the
$Z_8$ symmetry 
discussed in the previous subsection requires that
$\sum_{i}n_{i}$ and $\sum_{i}q_{i}$ are even for all
interactions, the theory is actually invariant under the
$Z_2$ transformation
\be
\label{KKparity}
\Upsilon^{(j,k)} (x^\mu) \mapsto (-1)^{j+k} \, \Upsilon^{(j,k)} (x^\mu) ~,
\ee
where $\Upsilon$ stands for either scalars or fermions.
Due to the dependence on KK numbers of the above transformation,
we refer to this symmetry as KK-parity, and we denote it by 
$Z_2^{\rm KK}$. 

We should note that operators localized at the points $(0,0)$ and
$(L,L)$ can potentially spoil this KK parity, unless they appear
symmetrically on the two branes.  The loop
induced localized operators automatically satisfy this requirement,
and therefore it is  natural for the above theories to contain the
KK-parity symmetry Eq.~(\ref{KKparity}), which implies that the
lightest KK mode, with  $(j,k) = (1,0)$,
is stable, and a good dark matter candidate 
\cite{Servant:2002aq, Cheng:2002ej}.

When fields with $l = 1$ are present, the folding boundary conditions
(\ref{summary-bc}) for $l=1$ imply that under the
reflection Eq.~(\ref{reflection}), fields of type $(n,l=1)$ are mapped
into fields of type $(n+2,l=1)$.  Explicitly, 
\be
f_{n,l=1}^{(j,k)}(\LL - x^4, \LL - x^5) = (-1)^{j+k+n} 
f_{n+2,l=1}^{(j,k)}(x^4, x^5)~.
\ee
Therefore, in order for the transformation (\ref{reflection}) to
be a symmetry of a theory containing $l=1$ fields, it is necessary
that both $(n,l=1)$ and $(n+2,l=1)$ fields be present.  We note,
however, that independently of whether the transformation
(\ref{reflection}) is a symmetry of the theory, when fields with $l=1$
are present the lightest KK mode is a $(j,k)=(1/2,1/2)$
state, which due to the $\Ztwol$ symmetry discussed in the previous
subsection [see Eqs.~(\ref{Z2symmetry}) and (\ref{Z2symmetryfermions})] 
is stable and can only be pair produced.

\section{Interactions of the Kaluza-Klein modes}

It is instructive to analyze how the symmetries discussed in the
previous section are realized in the KK picture.
To this end, one has to integrate products of KK wave-functions over the
square. 
Recall that the KK wave-functions for both scalars and fermions, 
$f_{n}^{(j,k)}(x^4, x^5)$, are given by Eq.~(\ref{KKsolns}),
with $j=0$ only if $n=k=0$, and $j>0$ otherwise.
These have the property
\be
\left(\partial_4 \pm i \partial_5 \right) f_{n}^{(j,k)}(x^4, x^5)
= \frac{i \pi }{L} (j \pm ik) f_{n \mp 1}^{(j,k)}(x^4, x^5) ~,
\ee
and therefore, whether or not an operator involves derivatives, 
its integral over the square is of the type
\be
\frac{1}{L^2}\int_{0}^{\LL} dx^4 \int_{0}^{\LL} dx^5 \,
f_{n_1}^{(j_1,k_1)} 
  ... f_{n_r}^{(j_r,k_r)}
= \frac{ 2^{2-r}  \Delta^{(j_1,k_1)...(j_r,k_r)}_{n_1 ...n_r} }{
	\left(1+\delta_{j_1, 0} \right) ...
  	\left(1+\delta_{j_r, 0} \right) }
\ee
where $r$ is the total number of fields that appear in the operator.
The $Z_8$ symmetry restricts the products of KK wave-functions 
to satisfy
\be
n_1+ ... + n_r = 0 \;{\rm mod}\, 4 ~.
\ee

In what follows we restrict attention only to fields with folding 
boundary conditions of the type $l=0$, for which $j$ and $k$ are
integers.
In order to compute $\Delta^{(j_1,k_1)...(j_r,k_r)}_{n_1 ...n_r}$
it is convenient to observe that the basic integral is
\be
I(j,k) \equiv \frac{1}{L^2}\int_{0}^{\LL} dx^4 \int_{0}^{\LL} dx^5
\,  e^{i\pi (j x^4 + k x^5)/L} ~,
\ee
which satisfies
\be
I(j,k) + I(-j,-k) + I(k,-j) + I(-k,j)  = 4 \delta_{j, 0}\delta_{k, 0} ~.
\ee
This property follows from the analytical continuation of the 
$I(j,k)$ integrals to the  $-L<x^{4},x^{5} <L$ region.

For a trilinear interaction ($r=3$) the result is
\bear
\Delta^{(j_1,k_1)(j_2,k_2)(j_3,k_3)}_{n_1,n_2,n_3}  \hspace*{-2.6em} &&
\hspace*{2.45em} =  \; 7 \delta_{j_1, 0}\delta_{j_2, 0}\delta_{j_3, 0} 
\nonumber \\ [0.6em]
&& \;	+  \delta_{j_1 + j_2, j_3}\delta_{k_1 + k_2, k_3} \, e^{i n_3\pi}
	+  \delta_{j_2 + j_3, j_1}\delta_{k_2 + k_3, k_1} \, e^{i n_1\pi}
	+ \delta_{j_3 + j_1, j_2}\delta_{k_3 + k_1, k_2} \, e^{i n_2\pi} 
\nonumber \\ [0.6em]
&& \;	+ \;\delta_{j_1 + k_2, j_3}\delta_{j_2 + k_3, k_1} \, e^{i (n_2/2 + n_3)\pi}
	+ \delta_{j_1 + k_3, j_2}\delta_{j_3 + k_2, k_1} \, e^{i (n_3/2 + n_2)\pi}
\nonumber \\ [0.6em]
&& \;	+ \;\delta_{j_2 + k_3, j_1}\delta_{j_3 + k_1, k_2} \, e^{i (n_3/2 + n_1)\pi}
	+ \delta_{j_2 + k_1, j_3}\delta_{j_1 + k_3, k_2} \, e^{i (n_1/2 + n_3)\pi}
\nonumber \\ [0.6em]
&&     \;   + \;\delta_{j_3 + k_2, j_1}\delta_{j_2 + k_1, k_3} \, e^{i (n_2/2 + n_1)\pi}
	+ \delta_{j_3 + k_1, j_2}\delta_{j_1 + k_2, k_3} \, e^{i (n_1/2 + n_2)\pi} ~.
\label{selection}
\eear
Note that the first term is nonzero only when all the KK modes are (0,0), while
the other nine terms are generic, describing interactions of various KK modes.
In general, for an interaction involving $r$ fields,
there are $t_r = (2^{r-1} - 1)^2$ generic terms and a term applying only to zero-modes
of coefficient $4^{r-1} - t_r$.

A particular case of interest for phenomenological applications
is the interaction of a number of zero modes with two higher modes. 
The integral over the square 
gives simply
\be
\frac{1}{L^2} \int_{0}^{\LL} dx^4 \int_{0}^{\LL} dx^5 \,
f_{n}^{(j_1,k_1)} f_{-n}^{(j_2,k_2)} f_{0}^{(0,0)} ...  f_{0}^{(0,0)}
= \delta_{j_1,j_2} \delta_{k_1,k_2} e^{-i n\pi} ~.
\ee

Eq.~(\ref{selection}) provides the tree-level selection rules for a trilinear interaction.
Given the KK numbers of two of the fields, $(j_1,k_1)$ and $(j_2,k_2)$,
the KK numbers of the third field are given by one of the following pairs:
\bear
\label{subset}
\hspace*{-1.6em}(j_3,k_3) & = & (j_1+j_2\, , \, k_1+k_2)\; , 
\; (j_1-j_2\, , \, k_1-k_2)\; , \; (-j_1+j_2\, , \, -k_1+k_2)\; ,  
\nonumber \\ [0.4em]
&&   (j_1+k_2\, , \, k_1-j_2)\; , \; (j_1-k_2\, , \, k_1+j_2)\; ,
\; (k_1+j_2\, , \, -j_1+k_2)\; , 
\nonumber \\ [0.4em]
&&   (-k_1+j_2\, , \, j_1+k_2)\; ,
\; (k_1-k_2\, , \, -j_1+j_2)\; , \; (-k_1+k_2\, , \, -j_1+j_2) ~.
\eear
All these choices are consistent with the KK parity discussed in the 
previous section, as they satisfy the condition that both $j_1+j_2+j_3$
and $k_1+k_2+k_3$ are even. However, not all values for $(j_3,k_3)$ 
allowed by KK parity are generated by the bulk interactions
at tree level. 
As in the five-dimensional 
case discussed in Ref.~\cite{Cheng:2002iz}, the other values for $(j_3,k_3)$ 
are generated by loops. For example, there is no tree-level coupling of 
two zero modes to a higher mode, while at one loop there is a coupling 
of two zero modes to the $(2j,2k)$, $(j+k, -j+k)$ and $(j-k, j+k)$
modes of $n=0$ fields, for any integer values of $j,k$, as 
shown in Figure 2.

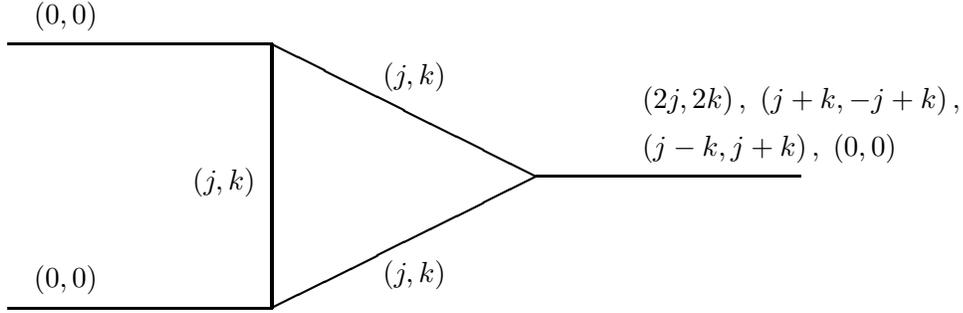
\begin{figure}
\begin{picture}(300,150)(-180,0)
\thicklines
\put(70,70){\line(1, 0){100}}
\put(70,70){\line(-2, 1){100}}
\put(70,70){\line(-2, -1){100}}
\put(-30,120){\line(-1, 0){100}}
\put(-30,20){\line(-1, 0){100}}
\put(-30,70){\line(0, 1){50}}
\put(-30,70){\line(0, -1){50}}
\put(110,96){$(2j,2k)\, ,\; (j+k, -j+k) \, ,$}
\put(110,78){$(j-k, j+k)\, ,\; (0,0)$}
\put(12,105){$(j,k)$}
\put(12,30){$(j,k)$}
\put(-60,65){$(j,k)$}
\put(-120,128){$(0,0)$}
\put(-120,28){$(0,0)$}
\end{picture}
%
\caption{Effective interaction of two zero modes with a higher KK mode.}
\label{schannelfig}
\end{figure}

\section{Connection to orbifold theories}

We now mention the relation between the compactification on the folded
square that we are studying and orbifold compactifications.  As has
become clear from the discussion in subsection 4.1 there is a close
connection between the theory on the folded square and rotations by
$\pi/2$ in an extended theory valid on the larger square $-\LL \leq
x^{4}, x^{5} \leq \LL$ (or even the full plane).
In fact, the theories we are studying can be obtained by starting from
fields defined on the whole six-dimensional space subject to periodic
or anti-periodic boundary conditions
\be
\label{periodic}
\Phi^{[p]}(x^{\mu},x^{4}+2\LL,x^{5}) = \Phi^{[p]}(x^{\mu},x^{4},x^{5}+2\LL)
= \Phi^{[p]}(x^{\mu},x^{4},x^{5})~,
\ee
or 
\be
\label{antiperiodic}
\Phi^{[a]}(x^{\mu},x^{4}+2\LL,x^{5}) = \Phi^{[a]}(x^{\mu},x^{4},x^{5}+2\LL)
= -\Phi^{[a]}(x^{\mu},x^{4},x^{5})~.
\ee
Fields that satisfy the periodic boundary condition are identified
with our $l=0$ fields, while those that satisfy the anti-periodic
one correspond to our $l=1$ fields.  In order  
to consistently impose the anti-periodic boundary condition,
Eq.~(\ref{antiperiodic}), it is necessary that the underlying
Lagrangian has a parity symmetry $\Phi^{[a]} \mapsto -\Phi^{[a]}$,
which corresponds to the $\Ztwol$ parity found in subsection 4.1
[see Eqs.~(\ref{Z2symmetry}) and (\ref{Z2symmetryfermions})].

Next, one can use the fact that the theory so defined has a symmetry
under rotations by $\pi/2$ about $(x^{4},x^{5})=(0,0)$ to perform a
further ``orbifold'' projection.  For scalars it reads
\be
\Phi(x^{\mu},-x^{5},x^{4}) = e^{i\theta} \Phi(x^{\mu},x^{4},x^{5})~,
\ee
where an internal $U(1)$ symmetry has been assumed that allows us to
identify fields at rotated points up to a phase.  For fermions, the
orbifold projection is
\be
e^{-i (\pi/2) \Sigma_{45}/2} \Psi(x^{\mu},-x^{5},x^{4}) = e^{i\theta} 
\Psi(x^{\mu},x^{4},x^{5})~.
\ee
We have explicitly shown in our derivation in sections 2 and 3 that
from all possible phases that are allowed a priori, only
those satisfying $\theta = n\pi/2$ with $n= 0,1,2,3$ 
allow for nontrivial
solutions.  The above comments further show that the compactification
on the folded square is equivalent to a $T^{2}/Z_{4}$ field-theory
orbifold, where the $Z_{4}$ corresponds to rotations by $\pi/2$ in the
plane of the compactified dimensions and the fields on the torus
$T^{2}$ may be periodic ($l=0$) or anti-periodic ($l=1$).
Consequently, the KK wave-functions for scalars\footnote{The KK wave-functions 
derived in Section 2 [see Eq.~(\ref{KKsolns})] agree with the ones given in 
Eq.~(3.12) of Ref.~\cite{Csaki:2002ur} except for a typo 
in that equation ($t^3 f_{p_y, -p_z}$ should read 
$t^3 f_{p_z, -p_y}$, and $t^3 f_{p,0}$ should read $t^3 f_{0,p}$).}
and fermions shown in Eqs.~(\ref{KKsolns}) and (\ref{KKsolns-fermion}), 
as well as the symmetries discussed in Section 4
are identical with the ones that can be derived by starting from 
the $T^{2}/Z_{4}$ compactification.

\section{Summary} \setcounter{equation}{0}

We conclude with a recapitulation of our main results 
regarding six-dimensional field theories with two 
dimensions compactified on a square. The boundary conditions 
prescribe the identification of two pairs of adjacent sides of the
square, and the equality of Lagrangian densities at 
identified points. 

A field has values at 
pairs of identified points which may differ 
by a symmetry transformation. 
In the case of a complex scalar
there is a $U(1)$ symmetry, so that the field values may differ
by a phase.
In general, the phase difference of a pair of adjacent sides
need not be equal to the phase difference of the other pair.
We refer to these as the ``folding" boundary conditions, 
and to the compactification in general as the ``folded square".
We have shown that
the compactification is smooth everywhere, 
in the sense that the derivative of the field in the
direction perpendicular to the identified boundaries is continuous
up to a phase.

The field equation has nontrivial solutions only if the 
phase that relates field values at identified points is 
$n\pi/2$, $n=0,1,2,3$. In addition, the phases
associated with the two pairs of 
identified sides may differ by $l\pi$, $l=0,1$. 
The most general folding boundary conditions  that 
allow solutions to the field equation for a complex scalar, 
labeled by the two integers $n$ and $l$, is given by Eq.~(\ref{summary-bc}).

It turns out that a folded square of size $L$ is equivalent 
to the compactification on a $T^2/Z_4$ orbifold where the 
$T^2$ is a torus of size $2L$. For a theory involving 
arbitrary interactions,
the $Z_4$ symmetry is a
requirement for the theory on the orbifold, whereas on the folded
square the same symmetry arises from the equality of the 
Lagrangian densities at identified points, combined with 
local six-dimensional Lorentz invariance.
The four values of $n$ are the possible values of the charge of 
the scalars under the $Z_4$ symmetry, while the two values of $l$ 
correspond to periodic ($l=0$) and anti-periodic ($l=1$) 
boundary conditions on $T^2$. The equality of the 
Lagrangian densities at identified points also implies 
the existence of a $Z_{2}$ symmetry, which we label $\Ztwol$, under which the fields have
charge $l$.

The wave functions of the KK modes, $f^{(j,k)}_n(x^4,x^5)$,
can be written as the sum of two cosines, as in Eq.~(\ref{KKsolns}). They
depend on two KK numbers, $j, k \ge 0$, which are integers
for $l=0$ and half-integers for $l=1$. 
Only the fields with $n=l=0$ have a zero mode, {\it i.e.}, $j=k=0$.
The completeness condition allows all the $k = 0$
states but none of the $j=0$ ones, 
except for the zero-mode.

Most of the above conclusions apply to fermions as well, with 
additional intricacies related to chirality.
In the case of a six-dimensional chiral fermion,   
the Dirac equation has non-trivial solutions only if
the folding boundary conditions for the left- and right-handed
four-dimensional chiralities are different. This is a key
property that allows the embedding of a four-dimensional 
chiral theory, such as the standard model, into a six-dimensional
theory with bulk fermions. 
Specifically, the left- and right-handed components of a 
fermion of six-dimensional chirality $\pm$ have folding boundary 
conditions with
values for $n$ that differ by $\pm 1$ mod 4. In particular, if the 
$+_L$ or $-_R$ ($-_L$ or $+_R$)
chirality has a zero mode, then the $+_R$ or $-_L$ ($+_L$ or $-_R$)
chirality has a wave function of type $n=3$ ($n=1$).
The phases of the KK wave functions
for the left- and right-handed components are also correlated,
their difference being given by the complex phase of 
$k\pm ij$ in the case of six-dimensional chirality $\pm $.

One difference compared to scalar theories is that in the presence of fermions
the $Z_4$ symmetry of the action is promoted to a $Z_8$ symmetry. This is a 
consequence of the six-dimensional Lorentz symmetry, or more precisely 
of the invariance under rotations by $\pi/2$ in the $(x^4, x^5)$ plane. 
In the context of the 
$T^2/Z_4$ orbifold \cite{Appelquist:2001mj}, the $Z_8$ symmetry is the group 
of rotations by $\pi/2$ around the center of $T^2$. 
From the point of view of the folded square,
$Z_8$ is an internal symmetry, with fermions carrying discrete charge
$q = \Sigma_{45}/2 + n$, where $\Sigma_{45}$ has eigenvalue $\mp 1$ for 
the $\pm_L$ chiralities, and $\pm 1$ for the $\pm_R$ chiralities.
For a scalar with folding boundary conditions of type 
$n$, the  $Z_8$ charge is $n$. The $Z_8$ symmetry requires that all operators 
in the four-dimensional effective theory
have a $Z_8$ charge given by 0 mod 4. In Ref.~\cite{Appelquist:2001mj}, 
it has been shown that 
for the standard model in six dimensions the $Z_8$ symmetry ensures a lifetime
for the proton longer than the current experimental bounds, even in  the
presence of baryon number violation at the TeV scale. Another implication
is that Majorana masses are forbidden (the implications for neutrino masses
are discussed in Ref.~\cite{Appelquist:2002ft}).
Although we restricted attention in this paper only to fermions and scalars,
the derivation of the $Z_8$ symmetry is based on general arguments
regarding the six-dimensional Lorentz symmetry, which hold in the presence 
of fields of any spin. The interesting case of gauge fields will be analyzed in
Ref.~\cite{Burdman}.

In theories where all fields satisfy boundary conditions with $l=0$, 
the above $\Ztwol$ symmetry becomes trivial. In this situation, however, 
the folded square compactification has one more symmetry, namely 
invariance under reflection with respect to the center of the 
square. 
In contrast to the above $Z_8\times \Ztwol$ symmetry, which assigns 
a unique charge to the whole tower of KK modes belonging to a 
six-dimensional field
(of given four-dimensional chirality, in the case of fermions),
the symmetry under reflection distinguishes between
KK modes. Therefore, this symmetry is a KK parity, and we label it by 
$Z_2^{\rm KK}$. 
A $(j,k)$ mode changes sign (remains invariant) under reflection
if $j+k$ is odd (even).
The $Z_2^{\rm KK}$ symmetry is similar to the KK parity of 
five-dimensional theories compactified on the $S^1/Z_2$
orbifold, which is invariance under reflection 
with respect to the center of the interval.
In six dimensions, however, reflection is part of the Lorentz 
symmetry, namely it is a rotation by $\pi$.
Therefore, any field theory on the folded square 
has an exact $Z_8\times Z_{2}$ symmetry, as a consequence 
of the equality of Lagrangian densities at identified points 
and of the six-dimensional Lorentz symmetry. If fields that 
satisfy boundary conditions with $l=1$ are present, 
then $Z_{2} = \Ztwol$. When all fields satisfy boundary 
conditions with $l=0$, then $Z_{2} = Z_2^{\rm KK}$. 
In either case, the $Z_{2}$ symmetry guarantees the 
stability of the lightest Kaluza-Klein mode, which 
may play the role of dark matter \cite{Servant:2002aq, Cheng:2002ej}.

The KK parity has important consequences on theories defined
on the folded square, since it provides a selection rule for 
any interaction: the sum of all 
KK numbers entering a vertex should be even.
The bulk interactions generate at tree level only a subset of these
vertices [see Eq.~(\ref{subset}) for the case of a trilinear interaction],
but quantum loops generate the other ones, similarly to the five-dimensional case
studied in \cite{Cheng:2002iz}.

In the compactification on the folded square, 
the corners of the square correspond to conical singularities. 
Two of the points are identified, while the other two
points are the fixed points of the $T^2/Z_4$ orbifold.
Typically, loops generate operators 
localized at these points with divergent coefficients. 
This suggests that physics at the cut-off scale could generate
additional contributions to the operators localized at these points. 
Nevertheless, such contributions are expected to be $Z_8$ 
invariant because the space in the vicinity of the corners has rotational 
symmetry. Moreover, as long as the underlying dynamics that induces
contributions at the cut-off scale does not distinguish between the 
fixed points,
the $Z_2^{\rm KK}$ symmetry is also preserved by any localized operators. 
When fields satisfying boundary conditions with $l=1$ are present, 
the $Z_2^{\rm KK}$ parity is generically broken by their boundary conditions, 
but a new $Z_{2}=\Ztwol$ symmetry emerges, as mentioned above. 

On the other hand, operators localized at the fixed points may perturb the 
KK spectrum.
At tree level, the squared KK masses are given by $(j^2 + k^2)(\pi /L)^2$ 
plus the squared mass of the six-dimensional field. Further contributions 
induced by loops include both finite pieces due to bulk kinetic terms 
and divergent pieces due to kinetic terms localized at the fixed points, 
as shown in 
\cite{Cheng:2002iz}. Likewise, contributions from physics above 
the cut-off scale
to kinetic terms localized at the fixed points would modify the KK masses.

The folded square is a simple 
compactification of two extra dimensions that allows 
the existence of chiral fermions in the four-dimensional 
effective theory.
Topologically, the boundary reduces to only a couple of points,
and therefore this compactification has good prospects for 
being the low energy behavior of some underlying dynamics.
Furthermore, the folded square has an intriguing symmetry structure.
It would therefore be interesting to compactify 
various six-dimensional extensions of the standard model
on the folded square.

\bigskip

\acknowledgments

We would like to thank Gustavo Burdman for collaboration 
during earlier stages of this work, and for many insightful 
conversations.
This work was supported by DOE under
contract DE-FG02-92ER-40704.

 \vfil \end{document}